\documentclass{JHEP3}

\usepackage{url,amsmath,amsthm,amscd,graphicx,psfrag}
\usepackage[all]{xy}

\def\tX{{\tilde X}}
\def\tV{{\tilde V}}
\def\pt{{\rm pt}}

\let\a=\alpha   \let\b=\beta        
              
   \let\l=\lambda  \let\m=\mu      
\let\n=\nu            \let\p=\pi      \let\r=\rho     \let\s=\sigma 
\let\t=\tau     \let\o=\omega            
               \let\S=\Sigma

   \def\cl{{\cal L}}
  \def\co{{\cal O}} \def\cp{{\cal P}}

 \def\IC{{\mathbb C}} \def\IP{{\mathbb P}}
   
\def\IZ{{\mathbb Z}} \def\IQ{{\mathbb Q}}

\theoremstyle{definition}

\newtheorem*{ackn}{Acknowledgments}

\theoremstyle{plain}

\theoremstyle{remark}

\def\plb#1 #2 {Phys. Lett. {\bf B#1} #2 }
\def\phr#1 #2 {Phys. Rep. {\bf  #1} #2 }    
\def\npb#1 #2 {Nucl. Phys. {\bf B#1} #2 }
\def\aph#1 #2 {Ann. Phys. {\bf #1} #2 }      
\def\jmp#1 #2 {J. Math. Phys. {\bf #1} #2 }
\def\jgp#1 #2 {J. Geom. Phys. {\bf #1} #2 }
\def\prd#1 #2 {Phys. Rev. {\bf D#1} #2 }
\def\prl#1 #2 {Phys. Rev. Lett. {\bf #1} #2 }
\def\rmp#1 #2 {Rev. Mod. Phys.  {\bf #1} #2 }
\def\zpc#1 {Z. Phys. {\bf #1C} }
\def\cmp#1 #2 {Commun. Math. Phys. {\bf #1} #2 }
\def\cqg#1 #2 {Class.Quant.Grav. {\bf #1} #2 }
\def\mpl#1 {Mod. Phys. Lett. {\bf A#1} }
\def\cpc#1 {Computer Phys. Commun. {\bf #1} }   
\def\ijmp#1 {Int. J. Mod. Phys. {\bf A#1} }
\def\ijmpC#1 {Int. J. Mod. Phys. {\bf C#1} }
\def\atmp#1 {Adv. Theor. Math. Phys. {\bf #1} }

\numberwithin{equation}{section}

\let\0=\over
\def\2{{1\over2}}
\let\6=\partial
\def\({\left(}       \def\){\right)}

\let\bra=\langle        \let\ket=\rangle        \def\<#1\>{\bra #1 \ket}
   \def\Im{\text{Im}}
\def\Aut{{\rm Aut}}

\newcommand{\be}{\begin{equation}}
\newcommand{\ee}{\end{equation}}
\title{On heterotic model constraints}

\author{Vincent Bouchard\\
Harvard University\\
Jefferson Physical Laboratory\\
17 Oxford St.\\
Cambridge, MA 02138, USA\\
E-mail: \email{bouchard@physics.harvard.edu}}

\author{Ron Donagi\\
Department of Mathematics\\
University of Pennsylvania\\
Philadelphia, PA 19104, USA\\
E-mail: \email{donagi@math.upenn.edu}}

\abstract{The constraints imposed on heterotic compactifications by global consistency and phenomenology seem to be very finely balanced. We show that weakening these constraints, as was proposed in some recent works, is likely to lead to frivolous results. In particular, we construct an infinite set of such frivolous models having precisely the massless spectrum of the MSSM and other quasi-realistic features. Only one model in this infinite collection (the one constructed in \cite{BD}) is globally consistent and supersymmetric. The others might be interpreted as being anomalous, or as non-supersymmetric models, or as local models that cannot be embedded in a global one. We also show that the strongly coupled model of \cite{BD} can be modified to a perturbative solution with stable $SU(4)$ or $SU(5)$ bundles in the hidden sector. We finally propose a detailed exploration of heterotic vacua involving bundles on Calabi-Yau threefolds with $\IZ_6$ Wilson lines; we obtain many more frivolous solutions, but none that are globally consistent and supersymmetric at the string scale.}

\begin{document}

\section{Introduction}

At low energies, string vacua are well approximated by effective field theories. 
However, not all effective field theories can be embedded in string theory. 
The string landscape is, presumably, a large but finite subset of the 
infinite landscape of all effective field theories. So in order that an effective field theory can be UV completed into a globally consistent string vacuum, it must satisfy some severe constraints. 
Our goal in this paper is to begin an exploration of some aspects of these constraints within the High Country region of the string landscape --- the region parametrizing those vacua that have the Standard Model spectrum and perhaps some further quasi-realistic properties. We restrict ourselves to a special class of string compactifications, namely compactifications of the $E_8 \times E_8$ heterotic string on non-simply connected Calabi-Yau threefolds. When the right constraints are satisfied, these compactifications provide string vacua with Standard Model features that can be constructed explicitly in a globally consistent way.

The constraints that need to be imposed are of two kinds: consistency requirements and phenomenological conditions. We review this in section \ref{s:Constraints}. There are two basic consistency requirements: a slope-polystability condition which is needed in order to have a solution of the Hermitian-Yang-Mills equations in the first place; and the Green-Schwarz anomaly cancellation condition. The precise phenomenological conditions we impose depend on how closely we want our model to approximate the Standard Model, as well as on the latest news from LHC. Here we simply require that the Standard Model group as well as a reasonable GUT group be obtained as the gauge groups on the Calabi-Yau and its universal cover, respectively; and that the low energy particle spectrum be that of the Standard Model (with no anti-generations or exotic particles, up to moduli fields). 
 
The High Country region is non empty. The only currently known vacuum with the above properties is the one constructed in \cite{BD}. This vacuum is briefly reviewed in section \ref{s:review}. 
It is a strongly coupled compactification of the $E_8 \times E_8$ heterotic string on a non-simply connected Calabi-Yau threefold with fundamental group $\IZ_2$. The compactification involves a visible stable $SU(5)$ bundle, which breaks the visible $E_8$ gauge group to an $SU(5)$ Grand Unified Theory (GUT). A $\IZ_2$ Wilson line is then used to break the $SU(5)$ to the MSSM gauge group $SU(3) \times SU(2) \times U(1)$. By computing various cohomology groups, it was shown that the massless spectrum of the low-energy effective theory consists in precisely the massless spectrum of the MSSM with no exotic particles, and with a choice of either $0$, $1$ or $2$ Higgs multiplets, depending on where one stands in the moduli space. Indeed, the model also includes $51$ moduli \cite{BCD}, which (one may hope) could be stabilized to realistic values by considering non-perturbative effects in the compactification. Beyond the explicit requirements above, this model also exhibits reasonable behavior of its trilinear couplings, which were computed at tree level in \cite{BCD}. It was shown that they are realistic enough not to rule out the model; some of them may even give mass to some of the neutrinos.

The size of the High Country region of the string landscape is not known. 
Given the immense size of the string landscape, one might expect that the High Country region too should be quite large. If not, one would like to understand why the constraints should turn out to be so unexpectedly strong. 
It is interesting that over two years after publication of \cite{BD}, that vacuum remains the only one known in this class of string compactifications. Various other models that satisfy some of these properties have however been constructed. One interesting class consists of heterotic orbifolds, as in \cite{BHLR}. Another model, considered in \cite{BHOP,BO}, will be discussed below. 

We would like to emphasize that the two consistency requirements and the basic phenomenological conditions are very finely balanced. Whether the High Country region turns out to be very large or very small, under any reasonable set of definitions it should turn out to be a finite set. One point we make in this paper is that any relaxation of the consistency requirements is liable to produce an infinite set of solutions. We are not saying that there should be an infinite landscape --- quite to the contrary, we are saying that any relaxation of the requirements that leads to such an infinite landscape must be physically suspicious. Let us, for example, frivolously remove the anomaly cancellation condition. In section 4 we construct an infinite family of such vacua satisfying all the remaining conditions; in particular they have precisely the massless spectrum of the MSSM and quasi-realistic tri-linear couplings. This infinite family is in fact obtained as a minor modification of the model of \cite{BD}. 

It is worth pointing out that within the heterotic theory this relaxation is precisely equivalent, as we recall in section \ref{s:consistency}, to admission of solutions involving a combination of $M5$- and anti-$M5$-branes, with supersymmetry broken at the compactification scale.
Indeed, introducing $M5$- and anti-$M5$-branes in the spectrum modifies the Green-Schwarz anomaly cancellation condition in such a way that for any desired visible bundle, the anomaly can be cancelled by introducing suitable $M5$- and anti-$M5$-branes. So our infinite family may alternatively be viewed as a family of non-supersymmetric vacua. From this point of view, finiteness can of course be restored if one takes into account that there is probably a maximal scale of supersymmetry breaking above which models become unstable. At the highest this would be the Planck scale, but possibly lower. The difficulty would then be converted into formulating explicit new constraints that guarantee an acceptable scale.\footnote{We thank Mike Douglas for correspondence on this issue. Similar comments on infinite families of non-supersymmetric vacua were made in \cite{AD}.}

A similar example occurs in \cite{BO}. This evolved in an attempt to restore consistency of the compactification of \cite{BHOP}. As was shown in \cite{GLS},
the compactification of \cite{BHOP} violates either polystability or anomaly cancellation, depending on the hidden sector bundle used; either way, it is not globally consistent. The proposal in \cite{BO} was to restore anomaly cancellation by adding to the compactification certain anti-$M5$-branes. The resulting vacuum is non-supersymmetric. The claim in \cite{BO} is that after moduli stabilization, some compactifications of the $E_8 \times E_8$ heterotic string on Calabi-Yau threefolds with both $M5$- and anti-$M5$-branes may be metastable, with long enough lifetime for them to be phenomenologically interesting \cite{BO, Bu, GLO}. It is certainly possible that this may apply to the bundle in \cite{BO}, and it is just as possible that it applies to any one of the infinite collection of bundles in our section 4.  It clearly should not apply to all, and with the present state of knowledge, we do not seem to have tools for deciding when it does and when it does not apply.

Another point of view on these vacua stems from the ``bottom-up" approach to string model building, whereby the idea is to focus on local properties of brane constructions, ignoring the global embedding in a fully consistent string theory. Dropping the topological anomaly cancellation may be understood as a bottom-up approach locally valid near the visible orbifold plane in heterotic M-theory. Note that this kind of bottom-up approach has been considered recently in detail from the F-theory dual point of view \cite{BHV}. It would be interesting to understand the analog of these infinite families on the F-theory side, perhaps using the dual description developed in \cite{CD, DW}, and to study more precisely how many of those F-theory local models can actually be completed to globally consistent string theories.

We are therefore left with the (so far unique) globally consistent Standard Model compactification of \cite{BD}. In section 3 we construct some variants of this model. The compactification of \cite{BD} was in the strong coupling regime of heterotic M-theory, since the presence of $M5$-branes was used to cancel the Green-Schwarz anomaly and produce a globally consistent compactification. 
Such non-perturbative compactifications have advantages as well as disadvantages for model building. Some of the advantages of perturbative heterotic compactifications, which do not have $M5$-branes, are discussed for instance in \cite{AC}.  So it is of interest to have a perturbative variant of our model.
Indeed, it turns out that the $M5$-branes in our model can be replaced by a polystable bundle in the hidden sector.
In this section we propose $SU(4)$ as well as $SU(5)$ hidden bundles which cancel the anomaly without introducing $M5$-branes. Hence, we obtain an alternative completion of the model of \cite{BD} to a fully pertubative and globally consistent heterotic compactification. We briefly comment on the possibility of using the resulting hidden theories to implement dynamical supersymmetry breaking, either through an analog of the Intriligator-Seiberg-Shih mechanism \cite{ISS} along the lines of \cite{BBO}, or through gaugino condensation.

To summarize the results discussed so far, we obtain three different string constructions with precisely the MSSM spectrum in the visible sector (up to moduli fields) and quasi-realistic tri-linear couplings, based on the model of \cite{BD}:
\begin{itemize}
\item A supersymmetric vacuum with $M5$-branes, in the strongly coupled regime of heterotic string theory, as presented in \cite{BD};
\item A supersymmetric vacuum without $M5$-branes, in the perturbative regime of heterotic string theory, with an $SU(4)$ or $SU(5)$ hidden theory which may perhaps be used to break supersymmetry dynamically (section 3);
\item An infinite family of (perhaps metastable) non-supersymmetric vacua with both $M5$- and anti-$M5$-branes (or bottom-up models near the visible orbifold plane), in the strongly coupled regime of heterotic string theory (section 4).
\end{itemize}

Finally, in section 5, we try to construct new quasi-realistic supersymmetric compactifications of string theory. In \cite{BD2}, a class of non-simply connected Calabi-Yau threefolds was constructed, by classifying all possible finite groups acting freely on smooth Calabi-Yau threefolds constructed as fiber products of two rational elliptic surfaces. The non-simply connected threefolds are then obtained by taking the quotients. This class of threefolds provides a playground to study how frequent quasi-realistic compactifications of heterotic string theory on non-simply connected Calabi-Yau threefolds are.

In particular, four distinct Calabi-Yau threefolds with fundamental group $\IZ_6$ were constructed. Here, we study existence of Standard Model bundles on these threefolds. A $\IZ_6$ Wilson line can be used to break either the $SU(5)$ GUT group to the MSSM gauge group $SU(3) \times SU(2) \times U(1)$, or the $Spin(10)$ GUT group to the MSSM gauge group with an extra $U(1)_{B-L}$. We propose various bundle constructions, but are unable to satisfy all the topological conditions required by global consistency and phenomenology, in the supersymmetric regime. The key point is that the two consistency conditions, namely the stability condition resulting from the Donaldson-Uhlenbeck-Yau theorem, and the anomaly cancellation condition with (or without) $M5$-branes in the spectrum, are very difficult to satisfy simultaneously. This is reminiscent of the analysis proposed in section 4, where only one model out of the infinite family actually satisfied both conditions. Here as well we obtain infinite families of models with stable bundles that could be used to construct non-supersymmetric vacua with $M5$- and anti-$M5$-branes, or bottom-up models, but none of those satisfy the anomaly cancellation condition without introducing anti-$M5$-branes. 

We should mention that we have not exhausted all possible bundle constructions on these $\IZ_6$ threefolds, and it is possible that other types of constructions may lead to Standard Model bundles. It would also be interesting to study other threefolds in the classification of \cite{BD2} along similar lines, since there is no reason \emph{a priori} to focus on the $\IZ_6$ or the $\IZ_2$ threefolds. We hope to report on that in future work.

\begin{ackn}
We would like to thank Mirjam Cveti\v c, Emanuel Diaconescu, Mike Douglas, Matthias Schuett, Alessandro Tomasiello and Martin Wijnholt for interesting discussions. We would especially like to thank Igor Dolgachev for very useful correspondence with respect to the calculations in the Appendix. R.D. is partially supported by NSF grant DMS 0612992, NSF Focused Research Grant DMS 0139799, and NSF Research and Training Grant DMS 0636606. V.B. would like to thank the Mathematics Department of University of Pennsylvania for hospitality while this work was initiated.
\end{ackn}

\section{Compactifications of the $E_8 \times E_8$ heterotic string}

\label{s:Constraints}

\subsection{Generalities}

We consider four-dimensional compactifications of $E_8 \times E_8$ heterotic string theory on Calabi-Yau threefolds. Such an heterotic vacuum is specified by a Calabi-Yau threefold $X$ and an holomorphic vector bundle $V_t \to X$, where $V_t$ splits into the direct sum $V_t = V \oplus V_h$; $V$ is the visible bundle with structure group $G \subseteq E_8$, and $V_h$ is the hidden bundle with structure group $G_h \subseteq E_8$. The gauge group of the resulting low-energy effective theory in the visible (resp. hidden) sector is given by the commutant $H$ of $G$ in $E_8$ (resp. $H_h$ of $G_h$ in $E_8$).

In the following we will choose the visible bundle $V$ to have structure group $G = SU(4)$ or $SU(5)$, such that its commutant is, respectively, $H = Spin(10)$ or $SU(5)$. Then, in order to obtain the MSSM gauge group, we choose $X$ to be non-simply connected, with fundamental group $\pi_1(X) = G_X$, where $G_X$ is a finite abelian group. This allows us to use a discrete $G_X$ Wilson line to break the GUT gauge group $H=SU(5)$ to the MSSM gauge group $SU(3) \times SU(2) \times U(1)$, or $Spin(10)$ to the MSSM gauge group with an extra $U(1)_{B-L}$.

\subsection{Consistency and phenomenological requirements}

\label{s:constraints}

There are two kinds of constraints that the vector bundle $V_t = V \oplus V_h$ must satisfy. The first kind consists in consistency constraints required by the heterotic string compactification; if the bundle fails to satisfy these constraints, it cannot be used to produce a globally consistent string vacuum. The second kind of constraints is phenomenological; they are necessary in order to obtain realistic four-dimensional physics.

\subsubsection{Consistency requirements}

\label{s:consistency}

Global consistency of the heterotic compactification imposes two particular constraints: 
\begin{itemize}
\item {\bf (S)}: At tree level, $V_t$ must be \emph{polystable} with respect to a certain K\"ahler class $\omega$ on $X$, and have zero slope $\m_\o(V_t) = 0$;
\item {\bf (A)}: $V_t$ must satisfy the Green-Schwarz \emph{anomaly cancellation condition}.
\end{itemize}

The first constraint comes from the Donalson-Uhlenbeck-Yau theorem, which states that a vector bundle admits an Hermitian Yang-Mills connection if and only if it is polystable with respect to the K\"ahler class. Let us now recall the definition of polystability, for completeness. Let $\omega$ be a K\"ahler class on $X$, and define the \emph{slope} $\mu_{\omega} (V_t)$ of a vector bundle $V_t$ on $X$, with respect to $\omega$, by
\be
\mu_{\omega}(V_t) = \frac{c_1(V_t) \cdot \omega^2}{\text{rank}(V_t)}.
\ee
A vector bundle $V_t$ is \emph{stable} with respect to $\omega$ if and only if 
\be
\mu_{\omega}(V') < \mu_{\omega} (V_t)
\ee
for all subbundles $V' \subset V_t$. It is \emph{semistable} if the inequality is weakened to
\be
\mu_{\omega}(V') \leq \mu_{\omega}(V_t).
\ee
$V_t$ is \emph{polystable} if it is a direct sum of stable bundles $V_t = V_1 \oplus \ldots \oplus V_k$ with equal slopes:
\be
\mu_{\omega}(V_t) = \mu_{\omega}(V_1) = \ldots = \mu_{\omega}(V_k).
\ee
Polystability, which is stronger than semistability but weaker than stability, is the condition required by the Donaldson-Uhlenbeck-Yau theorem, and relevant for global consistency of heterotic compactifications at tree level. In fact, at tree level consistency requires that the bundle $V_t$ satisfies the zero-slope limit of the Hermitian-Yang-Mills equation, which imposes that the slope of $V_t$ vanishes (see for instance \cite{We}). Note that we will impose further that $c_1(V_t) = c_1(V) = c_1(V_h) = 0$, which implies that $\m_\o (V_t) = 0$.

The second constraint comes from the Bianchi identity for the three-form field strength $H$ in heterotic string theory, which implies the topological constraint
\be
\label{e:ano}
c_2(TX) - c_2(V_t) = c_2(TX) - c_2(V) - c_2(V_h) =0
\ee
on the second Chern classes of the bundles $V$, $V_h$ and the threefold $X$. Note however that the Bianchi identity is in fact stronger than this topological condition, since it should be satisfied at the form level.\footnote{See for instance \cite{AC} for a recent discussion of this point.} However in the following we will be satisfied with requiring only the topological condition.

The anomaly cancellation condition \eqref{e:ano} must be satisfied for perturbative heterotic compactifications. In the strongly coupled regime of heterotic M-theory, it is possible to allow for the presence of $M5$-branes in the spectrum, which must wrap holomorphic curves in $X$. The anomaly cancellation condition then reads
\be\label{e:anom5}
c_2(TX) - c_2(V) - c_2(V_h) = [W],
\ee
where $[W]$ is the four-form Poincar\'e dual to the homology class of a (sum of) holomorphic curve(s) in $X$ around which the $M5$-branes are wrapped. Holomorphicity, which is required to preserve supersymmetry, translates into effectiveness of $[W]$. In terms of model building, this weakened condition means that we should look for bundles $V$ and $V_h$ such that the left-hand-side of \eqref{e:anom5} is effective, and cancel the anomaly by introducing $M5$-branes wrapping a (or some) curve(s) in this class.

Note that in \cite{BO} it was suggested that it is even possible to allow for the simultaneous presence of both $M5$-branes and anti-$M5$-branes in the spectrum, both wrapping holomorphic curves in $X$. Such configurations break supersymmetry explicitly at the compactification scale, and may potentially be physically unstable, although it was argued in \cite{BO} that they can be metastable with long enough lifetime. For these vacua the anomaly cancellation condition would become
\be
\label{e:anomalyanti}
c_2(TX) - c_2(V) - c_2(V_h) = [W] - [\bar W],
\ee
where $[\bar W]$ is now Poincar\'e dual to the homology class of a holomorphic curve in $X$ around which the anti-$M5$-branes are wrapped. Hence, both $[W]$ and $[\bar W]$ are effective, which means that the right hand side of \eqref{e:anomalyanti} is neither effective nor anti-effective. In terms of model building, this proposal has a dramatic consequence; namely, we can now look for any bundles $V$ and $V_h$ with no restriction whatsoever on their second Chern classes, and cancel the anomaly by choosing $[W]$ and $[\bar W]$ appropriately.

To summarize, the first string revolution gave us the exact condition \eqref{e:ano} on $c_2(V_t)$ for a perturbative compactification; the second string revolution enabled us to weaken this to an inequality \eqref{e:anom5} (the effectivity condition) on $c_2(V_t)$ for compactifications with $M5$-branes; while the work of \cite{BO} would give us blanket permission to waive the condition altogether for compactifications with both $M5$- and anti-$M5$-branes.

\subsubsection{Phenomenological constraints}

On top of these consistency conditions, there are more topological conditions that we require to obtain phenomenologically interesting physics. In the visible sector, we require that:
\begin{itemize}
\item {\bf (C1)}: $c_1(V) = 0$, since we want structure group $SU(n)$ rather than $U(n)$;\footnote{To be precise, this also comes from a global consistency requirement, namely that the second Stiefel-Whitney class of the bundle vanishes (see for instance \cite{We} for a recent discussion), which implies that $c_1(V) = 0 \text{ mod }2$. We require that $c_1(V) = 0$ since we want $SU(n)$ bundles; but recently the interesting possibility of constructing $U(n)$ bundles with $c_1(V) \neq 0$ has also been investigated, see for instance \cite{AHR}. Phenomenologically interesting $U(n)$ bundles have also been studied in \cite{BMRW,BMW,We}. Note that by requiring that $c_1(V) = c_1(V_h)=0$, we automatically satisfy the zero slope consistency condition on $V_t$.}
\item {\bf (C3)}: $c_3(V) = \pm 6$, to get three generations of particles.
\end{itemize}
Moreover, in the hidden sector, we require:
\begin{itemize}
\item {\bf (C1h)}: $c_1(V_h) = 0$, since we want structure group $SU(n)$ rather than $U(n)$;
\item {\bf (C3h)}: $c_3(V_h) = 0$, to have no chiral matter in the hidden sector.
\end{itemize}

Obviously, much more than these topological constraints must be satisfied in order to get realistic low-energy physics. For instance, in the visible sector, no exotic particle should appear in the massless spectrum, the Yukawa couplings should be realistic, the proton must not decay, etc. Some of these extra constraints can be studied as in \cite{BCD, BD,DHOR}, by computing cohomology groups of the bundle $V$ and triple products of cohomology groups. In the hidden sector, requiring that $c_3(V_h)=0$ does not mean that there is no generation/anti-generations pairs, but rather that the number of generations is equal to the number of anti-generations. To have no generation of particles at all in the hidden sector, we would need to strenghten the constraint by requiring that some cohomology groups of $V_h$ vanish, rather than just requiring that $c_3(V_h) = 0$.

Hence, the topological conditions above on the bundles are far from being sufficient for producing realistic low-energy physics, but are a first step --- which is already rather difficult to achieve --- towards this goal. Let us now describe the class of non-simply connected Calabi-Yau threefolds that we will be working with.

\subsection{The non-simply connected Calabi-Yau threefolds}
\label{e:fp}

We construct our non-simply connected Calabi-Yau threefolds by taking quotients of Schoen's Calabi-Yau threefolds $\tilde X = B \times_{\IP^1} B'$, which are smooth fiber products of two rational elliptic surfaces $\beta: B \to \IP^1$ and $\beta': B' \to \IP^1$ \cite{Sc}. $\tilde X$ can be represented by the commutative diagram
\begin{equation}\label{e:fibration}
\xymatrix{
& {\tilde X} \ar[dl]_{\pi'} \ar[dr]^{\pi} \\
B \ar[dr]^{\b} && B' \ar[dl]_{\b'} \\
& \IP^1
}
\end{equation}

In \cite{BD2} we classified all possible smooth fiber products $\tilde X$ admitting freely acting finite abelian groups of automorphisms $G_{\tilde X}$. By taking the quotients $X = \tilde X / G_{\tilde X}$, we obtained a large family of smooth non-simply connected Calabi-Yau threefolds $X$, with fundamental groups $\pi_1(X) = G_{\tilde X}$. These are the Calabi-Yau threefolds on which we will be compactifying heterotic string theory.

To be more precise, table 11 of \cite{BD2} presents the list of non-simply connected Calabi-Yau threefolds that can be constructed in this way. Each of these has one of the following fundamental groups:
\be
\IZ_3 \times \IZ_3, \quad \IZ_4 \times \IZ_2, \quad \IZ_6, \quad \IZ_5, \quad \IZ_4, \quad \IZ_2 \times \IZ_2, \quad \IZ_3, \quad \IZ_2.
\ee
In this paper we will focus on two particular classes of threefolds. First, we consider the non-simply connected Calabi-Yau threefold with fundamental group $\IZ_2$ corresponding to the next-to-last line in table 11 of \cite{BD2}. This is the Calabi-Yau threefold that was used in the heterotic standard model of \cite{BCD,BD}, and was first constructed in \cite{DOPW,DOPW2}. Its non-trivial Hodge numbers are $h^{1,1}(X) = h^{2,1}(X) = 11$. More precisely, in \cite{BCD,BD,DOPW,DOPW2} the covering Calabi-Yau threefold $\tilde X$, which admits a free $\IZ_2$ involution, was constructed explicitly as a smooth fiber product of two rational elliptic surfaces with configurations of singular fibers $\{ 2 I_2 , 8 I_1\}$; this is the generic configuration in the four-parameter family given by case $\#35$ in table 9 of \cite{BD2}.

The second class we will be interested in consists in the non-simply connected Calabi-Yau threefolds with fundamental group $\IZ_6$. According to table 11 of \cite{BD2}, there are four such threefolds, each of which has Hodge numbers $h^{1,1}(X) = h^{2,1}(X) = 3$. In section \ref{s:Z6} we make an attempt to construct Standard Model bundles on there manifolds. The reason why we chose this particular class of threefolds is that $\IZ_6$ turns out to be sufficiently big to break both $SU(5)$ and $Spin(10)$ to the MSSM gauge group (up to an extra $U(1)_{B-L}$ in the $Spin(10)$ case); hence we can try to construct both $SU(4)$ and $SU(5)$ visible bundles on these threefolds.

\section{Hidden bundles}

In this section we refine the heterotic compactification proposed in \cite{BD}, by constructing hidden bundles cancelling the anomaly, hence producing a fully perturbative compactification of heterotic string theory. It is conceivable that the resulting hidden theories may be used to implement dynamical supersymmetry breaking in the hidden sector. We briefly comment on this possibility, although we do not follow this line of research through in the present paper. But let us first review the construction of \cite{BD}.

\subsection{The heterotic standard model of \cite{BD}}
\label{s:review}

In this model we compactify the $E_8 \times E_8$ heterotic string theory on a Calabi-Yau threefold $X$ with $\IZ_2$ fundamental group. We construct a stable $SU(5)$ bundle on $X$ which is twisted by a $\IZ_2$ Wilson line to break the visible $E_8$ gauge group to the MSSM gauge group $SU(3) \times SU(2) \times U(1)$.

As mentioned earlier, the non-simply connected threefold $X$ is constructed by taking the quotient of a fiber product $\tilde X = B \times_{\IP^1} B'$ which admits a free $\IZ_2$ involution $\t : \tilde X \to \tilde X$; that is, $X = \tilde X / \langle \t \rangle$. We refer the reader to \cite{BD,BD2,DOPW} for the details of the geometry.

In the construction of \cite{BD}, we assumed the hidden bundle $V_h$ to be trivial --- this is the assumption we will want to relax later on. To get an $SU(5)$ visible bundle $V$ on $X$, we construct a $\IZ_2$-invariant bundle $\tilde V$ on the cover threefold $\tilde X$. The dual bundle $\tilde V^*$ is constructed as an extension
\begin{equation}
\label{e:extension}
0 \to V_2 \to \tV^* \to V_3 \to 0,
\end{equation}
where $V_2$ and $V_3$ are rank $2$ and $3$ bundles respectively given by
\begin{equation}
\label{e:vi}
V_i = \pi'^* W_i \otimes \pi^* L_i,
\end{equation}
where the $L_i$ are some line bundles on $B'$ and the $W_i$ are rank $i$ bundles on $B$ given by the Fourier-Mukai transforms $W_i = FM_B (C_i,N_i)$; as usual, the $C_i \subset B$ are curves in $B$ and the $N_i \in Pic(C_i)$ are line bundles over $C_i$.

Using the notation of section \ref{e:fp}, the explicit data goes as follows:
\begin{align}
\label{e:data}
{\bar C_2} &\in | \co_B ( 2 [e_0] +2[f])|,\notag \\
C_3 &\in | \co_B ( 3 [e_0] + 3 [f])|,\notag\\
C_2 &= {\bar C_2} + f_{\infty},\notag\\
N_2 &\in Pic^{3,1} (C_2),\notag\\
N_3 &\in Pic^7 (C_3),\notag\\
L_2 &= \co_{B'} ( 3 [r']),\notag\\
L_3 &= \co_{B'} ( -2 [r']).
\end{align}
$f_{\infty}$ is the smooth fiber of $\b$ at $\infty$, and $Pic^{3,1} (C_2)$ denotes line bundles of degree $3$ over $\bar C_2$ and degree $1$ on $f_{\infty}$. $[e_0]$ is the class of the zero section of $\b$, while $[f]$ is the fiber class. Finally, the class $[r']$ is given by
\begin{equation}
[r'] = [e_1'] + [e_4'] -[e_5'] + [e_0'] + [f'],
\end{equation}
where the $e_i'$'s are sections of $\b'$. Again, we refer the reader to \cite{BD, DOPW} for more details on this construction.

It was shown in \cite{BD} that $\tV$ is stable and invariant under the $\IZ_2$ involution.
Its cohomology was computed in \cite{BD}, and it leads to exactly the MSSM massless particle spectrum, with no exotic particles, up to moduli fields. The number of Higgs multiplet is either $0$, $1$ or $2$, depending on where we stand in the moduli space. It was also computed in \cite{BCD} that the spectrum contains $51$ vector bundle moduli, on top of the $11$ complex structure moduli and the $11$ K\"ahler moduli of the Calabi-Yau threefold $X$. Finally, the tri-linear couplings were computed at tree level in \cite{BCD}, and shown to be realistic enough not to rule out the model phenomenologically. 

It was also shown that the bundle satisfies the topological version of the anomaly cancellation condition. More precisely, we obtained that (recalling that the hidden bundle is assumed to be trivial)
\be
c_2(T \tX) - c_2(\tV) = 2 [f \times \pt] + 6 [\pt \times f'] := [W]. 
\ee
Hence, to make the vacuum globally consistent we must include $M5$-branes in the spectrum, wrapping a curve in the Poincar\'e dual of the effective class $[W]$. That is, our compactification is in the strongly coupled regime of heterotic M-theory.

The purpose of this section is to construct a $\IZ_2$-invariant hidden bundle $\tV_h$ on $\tX$ with second Chern class
\be\label{e:c2hid}
c_2(\tV_h) = 2 [f \times \pt] + 6 [\pt \times f'],
\ee
such that the anomaly cancellation condition is satisfied without the need for $M5$-branes, hence providing a fully perturbative compactification of the heterotic string.

\subsection{The construction of hidden bundles}

We now propose $SU(4)$ and $SU(5)$ hidden bundles which are $\IZ_2$-invariant, polystable, with second Chern class \eqref{e:c2hid}, and satisfying the topological conditions {\bf (C1h)} and {\bf (C3h)} presented earlier.

\subsubsection{An $SU(5)$ hidden bundle}

We first construct a hidden bundle $\tV_h$ as the direct sum of two $\IZ_2$-invariant stable bundles 
\be
\tV_h = \tV_1 \oplus \tV_2
\ee
 with $c_1(\tV_1) = c_1(\tV_2)=0$ --- hence $\tV_1$ and $\tV_2$ have zero slope. Then, $c_1(\tV_h) = 0$, and $\tV_h$ is polystable and $\IZ_2$-invariant. 

We construct $\tV_1$ by pulling back a rank $2$ bundle on $B'$, and $\tV_2$ by pulling back a rank $3$ bundle on the other rational elliptic surface $B$. That is,
\be
\tV_2 = {\pi'}^* W_2,\qquad \tV_1 = \pi^* W_1',
\ee
where $W_2$ lives on $B$, and $W_1'$ lives on $B'$.
$W_1'$ and $W_2$ are both constructed using a Fourier-Mukai transform. $W_1'$ is constructed from a pair $(C_1', N_1')$, where $C_1'$ is a curve in $B'$ and $N_1' \in Pic(C_1')$. We choose the data
\be
C_1' \in |\co_{B'}(2 [e_0'] + 2 [f']) |, \qquad N_1' \in Pic^{3}(C_1'),
\ee 
where $e_0', f'$ are respectively the zero section and the fiber of $B'$. Similarly, we construct $W_2$ on $B$ from the data
\be
C_2 \in |\co_{B}(3 [e_0] + 6 [f])|, \qquad N_2 \in Pic^{15} (C_2).
\ee
According to \cite{DOPW2}, both $C_1'$ and $C_2$ can be chosen to be smooth and $\IZ_2$-invariant, and the line bundles can also be chosen to be invariant under the Fourier-Mukai transformed involution. Hence the bundles $W_1'$ and $W_2$ are stable and $\IZ_2$-invariant.

Now it is easy to compute the Chern classes of $\tV_h$ from this data. Using the results of \cite{DOPW2}, we obtain
\be\label{e:chernhid}
c_1(\tV_1) = c_1(\tV_2) = c_1(\tV_h)=0, \qquad c_3(\tV_h)=0,\qquad c_2(\tV_h)= 2 [ f \times pt]+ 6 [pt \times f'].
\ee
So this bundle satisfies the conditions above on the Chern classes. Hence, we constructed an $SU(5)$ hidden bundle that cancels the anomaly, is polystable, is $\IZ_2$-invariant and satisfies the topological conditions on the Chern classes. 

\subsubsection{An $SU(4)$ hidden bundle}

The second construction is very similar to the first. We now construct $\tV_h$ as a direct sum of two rank $2$ stable bundles
\be
\tV_h = \tV_1 \oplus \tV_2
\ee
 with $c_1(\tV_1) = c_1(\tV_2)=0$. The bundles are constructed as
\be
\tV_2 = {\pi'}^* W_2,\qquad \tV_1 = \pi^* W_1',
\ee
where $W_2$ and $W_1'$ are constructed via Fourier-Mukai transforms with data
\be
C_1' \in |\co_{B'}(2 [e_0'] + 2 [f']) |, \qquad N_1' \in Pic^{3}(C_1'),
\ee 
and
\be
C_2 \in |\co_{B}(2 [e_0] + 6 [f])|, \qquad N_2 \in Pic^{11} (C_2).
\ee
Again, according to \cite{DOPW2}, both $C_1'$ and $C_2$ can be chosen to be smooth and $\IZ_2$-invariant, and the line bundles can be chosen to be invariant under the Fourier-Mukai transformed involution. Hence the bundles $W_1'$ and $W_2$ are stable and $\IZ_2$-invariant.

Computing the Chern classes, we get again \eqref{e:chernhid}. Thus, we constructed an $SU(4)$ hidden bundle that cancels the anomaly, is polystable, is $\IZ_2$-invariant and satisfies the topological conditions on the Chern classes.

\subsection{Dynamical supersymmetry breaking in the hidden sector}

In the previous section, we introduced non-trivial $SU(4)$ and $SU(5)$ hidden bundles in the compactification, which produce respectively hidden $Spin(10)$ and $SU(5)$ gauge theories. From a phenomenological point of view, it would be interesting if the hidden theory could be used to implement dynamical supersymmetry breaking. 

One approach that was explored in \cite{BBO} is to construct hidden theories with non-supersymmetric metastable vacua, providing a supersymmetry breaking mechanism that could, in principle, be mediated to the visible sector via standard mechanisms. Non-supersymmetric metastable vacua in gauge theories have been studied by Intriligator, Seiberg and Shih in \cite{ISS}. Roughly speaking, they considered $N=1$ supersymmetric $SU(N_c)$ theory with $N_f$ massive flavors in the fundamental representation. Then, they showed that if 
\be
N_c + 1 \leq N_f < \frac{3}{2} N_c,
\label{e:cond1}
\ee
there exists supersymmetry breaking vacua. Moreover, if 
\be\label{e:longlived}
|\epsilon| = \sqrt{\left | \frac{m}{\Lambda} \right| } \ll 1,
\ee
where $m$ is the typical scale of the quark masses and $\Lambda$ is the strong coupling scale, these vacua are long-lived.

They also considered $N=1$ supersymmetric $Spin(N_c)$ theories with $N_f$ massive flavors in the fundamental representation. In this case, supersymmetry breaking metastable vacua exist if
\be
N_c - 4 \leq N_f \leq \frac{3}{2} (N_c - 2),
\label{e:cond2}
\ee
and they are long-lived if the inequality \eqref{e:longlived} is satisfied.

One could hope that the $SU(5)$ and $Spin(10)$ hidden theories that we introduced in the previous subsection are of this type. We constructed $SU(4)$ and $SU(5)$ hidden bundles $V_h$ satisfying the condition $c_1(V_h) = c_3(V_h) = 0$, which means that the number of generations is equal to the number of anti-generations in the hidden sector. But it does not mean that these numbers are zero individually; for this, we need to show that $h^1(X,V_h) = h^1(X,V^*_h) = 0$. 

In fact, it is easy to compute that for both hidden bundles that we constructed above, $h^1(X,V_h) = h^1(X,V^*_h) \neq 0$. Hence our hidden theories have massless matter in the ${\bf 16}$ and ${\bf \overline{16}}$ of $Spin(10)$, and the ${\bf 10}$ and ${\bf \overline{10}}$ of $SU(5)$, respectively, and are not of the type studied by Intriligator, Seiberg and Shih in \cite{ISS}, which only have matter in the fundamental representation. However, the hidden theories that we constructed may still have non-supersymmetric metastable vacua; it would be interesting to study this question further. 
It would also be interesting to try to construct other hidden bundles which cancel the anomaly of our visible bundle, and produce hidden theories precisely of the type studied in \cite{ISS}. 

Another, more traditional, approach to supersymmetry breaking is to consider confining hidden theories, such that gaugino condensation produces an effective superpotential. Minimization of the potential may lead to vacua with supersymmetry dynamically broken (see for instance \cite{CLW} for such scenario in type II theories with intersecting $D6$-branes). It is possible that the hidden theories we constructed are confining; one needs to check whether the beta function is negative, which necessitates the computation of the number of particles in the fundamental representation. If this is the case, gaugino condensation should produce an effective superpotential; minimization of the potential could then lead to vacua with supersymmetry dynamically broken. It would also be interesting to construct hidden theories with at least two confining gauge group factors, which may implement the so-called racetrack mechanism for supersymmetry breaking. 

We postpone these investigations of supersymmetry breaking for future work.

\section{An infinite family of frivolous models}

\subsection{Motivation}

In this section we address the proposal of \cite{BO} to consider non-supersymmetric vacua of the strongly coupled heterotic string, with both $M5$-branes and anti-$M5$-branes in the spectrum. Let us be a little more precise. 

The $E_8 \times E_8$ heterotic string theory compactified on a Calabi-Yau threefold $X$ can be understood as $11$-dimensional heterotic M-theory compactified on the manifold
$
X \times S^1 / \IZ_2.
$ 
At each end of the interval $S^1 / \IZ_2$, there is an orbifold plane, on which an $E_8$ gauge theory lives; one end is chosen to be the visible sector, and the other end is the hidden sector. The $E_8$ visible and hidden gauge theories may be broken by introducing non-trivial stable vector bundles $V$ and $V_h$ on the orbifold planes. Moreover, one can introduce $M5$- and anti-$M5$-branes in the bulk space between the orbifold planes, wrapping holomorphic curves in the Calabi-Yau threefold $X$. Supersymmetry is then broken explicitly at the compactification scale.\footnote{Note that if no anti-$M5$-branes are present, supersymmetry remains unbroken.}

Supersymmetry is usually required to ensure physical stability of the vacuum. However, it was argued in \cite{BO, Bu}, and then in \cite{GLO}, following the pioneer work of KKLT in type II theory \cite{KKLT}, that after moduli stabilization, some heterotic vacua with both $M5$- and anti-$M5$-branes may be metastable, with long enough lifetime to be phenomenologically interesting, and with a small positive cosmological constant. Due to the present state of knowledge on moduli stabilization in heterotic string theory, it remains unclear to us which precise conditions should be satisfied for heterotic compactifications with $M5$-branes and anti-$M5$-branes to have metastable vacua with long enough lifetime after moduli stabilization. While it would be very interesting to understand these issues better, in this paper we will not address any of these claims explicitly, and content ourselves with a more pragmatic approach consisting in studying the effect of allowing such vacua on the mathematics of heterotic model building.

From the point of view of model building, the main consequence of the introduction of $M5$- and anti-$M5$-branes in the spectrum is to modify the anomaly cancellation condition, as explained in section \ref{s:constraints}. It considerably simplifies the mathematics, since for any bundles $V$ and $V_h$ the anomaly can be cancelled by introducing suitable $M5$- and anti-$M5$-branes. Indeed, one of the motivations behind the proposal of \cite{BO} was to make the compactification of \cite{BHOP} globally consistent, since without the introduction of anti-$M5$-branes in the spectrum it is anomalous (\emph{i.e.} it does not satisfy the condition \eqref{e:anom5}).

\subsection{The infinite family}

We now consider the heterotic Standard Model of \cite{BD}, which was reviewed in section \ref{s:review}. As in \cite{BD} we assume the hidden bundle to be trivial. What we now show is that if we drop the anomaly cancellation condition and allow $c_2(TX) - c_2(V)$ to be non-effective as in \eqref{e:anomalyanti}, we get an infinite family of heterotic string vacua with exactly the MSSM massless spectrum with no exotic particles up to moduli fields, precisely as in \cite{BD}, and quasi-realistic tri-linear couplings at the classical level as in \cite{BCD}.

We take the same threefold as in section \ref{s:review}, and construct the visible bundle $\tV$ by the extension \eqref{e:extension}, with $V_2$ and $V_3$ as in \eqref{e:vi}. Moreover, we use the same data as in \eqref{e:data}, except for the definition of the line bundles $N_2$ and $N_3$. We generalize the construction of the model by defining the line bundles to be
\be
N_2 \in Pic^{d_2-1,1} (C_2),\qquad N_3 \in Pic^{d_3} (C_3)
\ee
for any integers $d_2,d_3 \in \IZ$. The construction of \cite{BD} then consists in the particular case $(d_2,d_3) = (4,7)$. It is clear from the data \eqref{e:data} and the analysis of \cite{DOPW2,BD} that such bundles are $\IZ_2$-invariant.

Let us now compute the Chern character of this visible bundle $\tV$. Using the results of \cite{DOPW2}, we get that
\begin{equation}
ch(\tV)= 5 + (11-d_2-d_3) [\pi^* f'] - \big(6 [ \pt \times f'] + (-6 d_2 + 4 d_3 + 6) [f \times \pt] \big)  + 6 [\pt] .
\label{e:cc}
\end{equation} 
with the notation as in section \ref{s:review}. 
We find that $c_3(\tV)=12$ as required, while
 $c_1 (\tV) = 0 $ if 
\be
d_2 = 4-d, \qquad d_3 = 7 +d,
\ee
for some $d \in \IZ$.

If we require that the anomaly cancellation condition \eqref{e:anom5} be satisfied, \emph{i.e.} that $c_2 (\tX) - c_2(\tV)$ be effective, with $c_2(\tX) = 12 [\pt \times f'] + 12 [f \times \pt]$, we obtain the extra constraint
\begin{equation}\label{e:anom}
-6d_2 +4 d_3 \leq 6,
\end{equation}
that is
\be
d \leq 0.
\ee

As explained in \cite{DOPW2}, stability of $\tV$ requires that $H^1(\tX, V_2 \otimes V_3^*) \neq 0$ --- or, in other words, that there exists non-trivial extensions of the form \eqref{e:extension} --- and that $c_1(V_2)$ be non-effective. From \cite{DOPW2,BD}, we know that the first condition is automatically satisfied since $L_2 \cdot f' > L_3 \cdot f'$. The first Chern class of $V_2$ is given by
\be
c_1(V_2) = 6 [\pi^* r'] - (d+1)[\pi^* f'];
\ee
one can show that the second condition imposes that
\begin{equation}\label{e:stab}
d \geq 0.
\end{equation} 
To summarize, if we do not require anomaly cancellation, the integer $d \in \IZ$ must only satisfy $d \geq 0$, for stability.
Anomaly cancellation adds the extra constraint that $d \leq 0$, therefore the only solution to both constraints is $d=0$.

What we have just found is that if we do not require the anomaly cancellation constraint \eqref{e:anom}, we can choose any integer $d \geq 0$ to get a visible bundle satisfying the stability condition, $\IZ_2$-invariance, and the phenomenological requirements on the Chern classes. Moreover, it is easy to see that all the cohomology calculations to determine the massless spectrum in \cite{BD}, and similarly for the triple product calculations to obtain the tri-linear couplings in \cite{BCD}, {\em do not} depend on the degrees of the line bundles $d_2$ and $d_3$, hence on $d$. Hence, all the results of \cite{BCD,BD} carry over to these new constructions.

As a result, we get an infinite family of models, parameterized by $d \in \IZ$, $d \geq 0$, with exactly the massless spectrum of the MSSM with no exotic particles, and quasi-realistic tri-linear couplings at tree level. Note that all these models are really different, since varying the integer $d$ changes the second Chern class of the bundle $\tV$, which reads (see \eqref{e:cc})
\be
c_2(\tV) = 6 [ \pt \times f'] + 10(d+1) [f \times \pt].
\ee 
Of course, all of these models are non-supersymmetric (except for $d=0$ where we recover the model of \cite{BD}), in the sense that they do not satisfy the anomaly cancellation condition \eqref{e:anom} unless one introduces both $M5$- and anti-$M5$-branes in the spectrum.

To summarize, in the heterotic context with unbroken supersymmetry, there seems to be a strong tension between the two consistency conditions of the compactification; namely, the stability condition on the bundles coming from the Donaldson-Uhlenbeck-Yau theorem, see \eqref{e:stab}, and the topological version of the anomaly cancellation condition, see \eqref{e:anom}. If we drop the anomaly cancellation condition, instead of a unique solution we get an infinite family of solutions, which can be interpreted as non-supersymmetric solutions with $M5$- and anti-$M5$-branes. This also makes sense from the local point of view; without requiring anomaly cancellation, we get an infinite family of quasi-realistic supersymmetric effective theories, but only one of them can be UV completed into a globally consistent string theory. This shows explicitly that given a local compactification, it may be very difficult, if not impossible, to find a global embedding in a consistent supersymmetric string theory.

In fact, we will see the same tension between these consistency conditions in the next section, when we try to construct Standard Model bundles on another class of threefolds. We will also get infinite families of non-supersymmetric vacua, but in this case though, we will not be able to find any solution to the two constraints; that is, none of these local models will have a consistent global supersymmetric string completion.

\section{A study of $\IZ_6$ models}
\label{s:Z6}

In this section, we try to construct new Standard Model bundles on a class of non-simply connected Calabi-Yau threefolds with $\IZ_6$ fundamental group.

\subsection{The Calabi-Yau threefolds}

We start again with Schoen's Calabi-Yau threefolds $\tX = B \times_{\IP^1} B'$, which are fiber products of two rational elliptic surfaces. We want to construct smooth fiber products $\tX$ that admit a free $\IZ_6$ group of automorphisms $\langle \tau \rangle$, where $\tau: \tX \to \tX$ is an order 6 automorphism, and take the quotient to obtain non-simply connected Calabi-Yau threefolds $X = \tX / \langle \tau \rangle$.

We classified fiber products $\tX$ with finite order group of automorphisms in \cite{BD2}. We found four different non-simply connected Calabi-Yau threefolds $X$ with $\IZ_6$ fundamental group (see table 11 of \cite{BD2}). To understand the differences between the four threefolds, we need to consider the $\IZ_6$ automorphism group a little closer.

Any automorphism $\tau : \tX \to \tX$ that preserves the elliptic fibration is a fiber product 
\be
\t = \t_B \times_{\IP^1} \t_{B'}
\ee
 of two automorphisms $\t_B: B \to B$ and $\t_{B'}: B' \to B'$ of the rational elliptic surfaces. For the group $\langle \t \rangle$ to be free, the groups $\langle \t_B \rangle$, $\langle \t_{B'} \rangle$ and $\langle \t \rangle$ must all be equal abstractly. Following the fibration structure \eqref{e:fibration}, $\tau$ induces an automorphism $\t_{\IP^1}: \IP^1 \to \IP^1$ of the $\IP^1$-base, which is given by its projection:
\be
\t_{\IP^1} \circ \b = \b \circ \t_B, \qquad \t_{\IP^1} \circ \b' =  \b' \circ \t_{B'}.
\ee
Generically, $\langle \t_{\IP^1} \rangle$ will only be a sugroup of $\langle \tau \rangle$, since some of the elements of $\langle \tau \rangle$ may act trivially on the $\IP^1$ (for instance if they consist of only translation by torsion sections on $B$ and $B'$).

To construct fiber products $\tX$ with particular finite order group of automorphisms, one needs to consider non-generic rational elliptic surfaces $B$ and $B'$, with specific configurations of singular fibers. Let us now focus on the case where $\tau$ is an order 6 automorphism. According to our classification in \cite{BD2}, it turns out that there are four possible ways of combining non-generic rational elliptic surfaces $B$ and $B'$ such that the fiber product $\tX$ admits a free $\IZ_6$ action. The four families can be differentiated by the induced automorphism on $\IP^1$. Namely, the four families have, respectively, 
\be
\langle \t_{\IP^1} \rangle \simeq 1, \IZ_2, \IZ_3, \IZ_6,
\ee 
which are the four possible subgroups of $\langle \t \rangle \simeq \IZ_6$. By taking the quotients $X = \tX / \langle \t \rangle$, we thus obtain four non-simply connected Calabi-Yau threefolds $X$ with $\pi_1(X) = \IZ_6$.

These threefolds were studied in detail in \cite{BD2}. In particular, it was proved that the non-trivial Hodge numbers of the four threefolds are $h^{1,1}(X) = h^{2,1}(X) = 3$. We refer the reader to \cite{BD2} for the details of the geometry. In this paper what will be required for the construction of the bundles is a better understanding of the invariant cohomology of $\tX$, and its intersection ring, to which we now turn to.

\subsection{The invariant cohomology of $\tX$}

\label{s:cohom}

In the following we will attempt to construct Standard Model bundles on the non-simply connected Calabi-Yau threefolds $X$ with $\pi_1(X) = \IZ_6$. More precisely, we will construct $\langle \tau \rangle$-invariant bundles on the cover threefolds $\tX$, which descend to bundles on the quotient threefolds. To this end, we need to understand the invariant cohomology of $\tX$. In fact, as will become clear as we go along, what we want to compute is the invariant cohomology subgroup of $H^2(B, \IZ)$ under the cohomological action of $\langle \tau_B \rangle$ --- and similarly for $B'$. We denote the invariant subgroups with a superscript, such as $H^2(B,\IZ)^{\langle \t_B \rangle}$.

For the Calabi-Yau threefolds $X$ with $\pi_1(X)$, we showed in \cite{BD2} that
\be
\dim H^2(X, \IZ) = \dim H^2(\tX, \IZ)^{\langle \t \rangle} = 3.
\ee
From the fiber product structure of $\tX$, we have that
\be
\dim H^2(\tX, \IZ)^{\langle \t \rangle} = \dim H^2(B, \IZ)^{\bra \t_B \ket} + \dim H^2(B', \IZ)^{\bra \t_{B'} \ket}-\dim H^2( \IP^1, \IZ).
\ee
Hence, we obtain
\be
\dim H^2(B, \IZ)^{\bra \t_B \ket} = \dim H^2(B', \IZ)^{\bra \t_{B'} \ket} = 2.
\ee
That is, each rational elliptic surface has a two-dimensional invariant cohomology group.

Let us now focus on the rational elliptic surface $B$ and its automorphism group $\bra \t_{B} \ket$ --- the same holds for $B'$. We want to find the two invariant classes generating $H^2(B, \IZ)^{\bra \t_B \ket}$, and their intersection numbers. First, any automorphism must preserve the canonical class, which in the case of a rational elliptic surface is just $K_B = -[f]$, where $[f]$ is the fiber class of the elliptic fibration. Hence, $[f]$ is an invariant class under $\bra \t_{B} \ket$. By definition, its self-intersection is $[f] \cdot [f] = 0$.

The construction of the second invariant class is slightly more involved. Take the generator $\t_B$ of the automorphism group. In full generality (see \cite{BD2}), it can be written as
\be
\label{e:linear}
\t_B = t_{\xi} \circ \a_B,
\ee
where $\a_B$ is the linearization of $\t_B$ which fixes the zero section $\s$ of the elliptic fibration, and $t_{\xi}$ denotes translation by a section $\xi \in MW(B)$ in the Mordell-Weil group of $B$. It was shown that the order of $\alpha_B$ is equal to the order of the projection $\t_{\IP^1}$. Hence, for our four families we get that $\bra \a_B \ket \simeq 1, \IZ_2, \IZ_3, \IZ_6$ respectively.

Now consider the zero section $\s: \IP^1 \to B$, which defines a class $[\s] \in H^2(B,\IZ)$. The automorphism $\t_B$ induces an automorphism on cohomology, which we denote by $\t_B^*$. Apply $\t_B^*$ six times to the class $[\s]$, and take the sum:
\be
[\l]:= \sum_{i=0}^5 (\t_B^*)^i [\s].
\ee
Since $(\t_B^*)^6 = 1$, we get that $\t_B^* [\l] = [\l]$, hence $[\l]$ is an invariant class.

Moreover, from \eqref{e:linear} we see that $\t_B^*B [\s] = [\xi]$, which is just the class of the section $\xi$. Similarly 
\be
(\t_B^*)^i [\s] = [e_i] := [\a_B^{i-1} \xi + \a_B^{i-2} \xi + \ldots + \xi], \qquad i=1, \ldots, 5,
\ee
where we defined the sections $e_i \in MW(B)$, $i=1,\ldots,5$. If we finally denote the zero section $\s$ by $e_0$, we get that
\be
[\l] = \sum_{i=0}^5 [e_i],
\ee
which is just the sum of the cohomology classes of six sections. Hence, we obtain directly the intersection product $[\l] \cdot [f] = 6$.

The remaining intersection product $[\l] \cdot [\l]$ is not fixed from general arguments, since it depends on whether the sections $e_i$ intersect or not, which in turn depends on the choice of $\xi$ in the definition of the automorphism \eqref{e:linear}. We postpone the calculation of the self-intersection of $[\l]$ to the Appendix, since it is rather technical. Let us simply state here the result. For all four families, and all allowed choices of $\xi$, we always obtain
\be
[\l] \cdot [\l] = -6.
\ee
In other words, we can assume, in full generality, that the six sections $e_i$, $i=0,\ldots,5$, are mutually disjoint.

To summarize, we have constructed two $\IZ_6$-invariant classes $[f],[\l] \in H^2(B,\IZ)$ for any of the four Calabi-Yau threefolds, with intersection numbers
\be
\label{e:intfull}
[f] \cdot [f] = 0, \qquad [f] \cdot [\l] = 6, \qquad [\l] \cdot [\l] = -6.
\ee
However, we have not showed yet whether these two classes generate the full invariant cohomology group $H^2(B, \IZ)^{\bra \t_B \ket}$, or just a finite index subgroup thereof. We again postpone the analysis to the Appendix, the result being that $[f]$ and $[\l]$ indeed generate the full invariant cohomology group $H^2(B, \IZ)^{\bra \t_B \ket}$.

\subsection{$\IZ_6$ symmetry breaking pattern}

Before we turn to the construction of bundles, let us explain in more detail why we decided to concentrate on the threefolds with $\IZ_6$ fundamental group. The main reason is that $\IZ_6$ can be used to break both $SU(5)$ to the MSSM gauge group $SU(3) \times SU(2) \times U(1)$, and $Spin(10)$ to the MSSM gauge group with an extra $U(1)_{B-L}$. Hence, we have the extra freedom of trying to construct both $SU(5)$ and $SU(4)$ bundles on $\tX$. Here for completeness we describe explicitly the two possible symmetry breaking patterns.

To break $SU(5)$ to the MSSM gauge group, we embed the non trivial element $-1$ of $\IZ_2$ into
$SU(5)$ diagonally as 
\be
\text{diag}(1,1,1,-1,-1).
\ee %(This is known in physics as the Georgi-Glashow embedding.) 
The commutator contains $SU(3) \times SU(2)$ embedded in the upper left and bottom right corners, respectively, plus a
$U(1)$ parametrizing diagonal matrices of the form
\be
\text{diag}(\lambda^2,\lambda^2,\lambda^2,\lambda^{-3},\lambda^{-3}),\qquad \lambda \in U(1).
\ee
The actual commutator group is $(SU(3) \times SU(2) \times U(1))/ \IZ_6$, where the
denominator $\IZ_6$ is the intersection in $SU(5)$ of $SU(3) \times SU(2)$ with $U(1)$.

Let us now turn to $Spin(10)$. The maximal torus of $SO(10)$ can be naturally identified with $(SO(2))^5$.\footnote{If the quadratic form preserved by $SO(10)$ is written as
$\sum_{i=1}^5 x_i y_i$,
then the five $SO(2)$'s operate on one $(x_i, y_i)$ pair at a time.} 
The maximal torus for $Spin(10)$ is a double cover which induces a non-trivial double
cover of each $SO(2)$. An element of order $2$, such as:
\be
(1,1,1,1,-1) \in (SO(2))^5 \subset Spin(10),
\ee
corresponding to a $180^o$ rotation, thus lifts to an element of order $4$ in
$Spin(10)$. But the product of two such, for example
\be
(1,1,1,-1,-1) \in (SO(2))^5 \subset Spin(10),
\ee
lifts to an element $\alpha$ of order $2$ in $Spin(10)$. Its commutator in
$Spin(10)$ is
\be
Spin(6) \times Spin(4) = SU(4) \times SU(2) \times SU(2).
\ee
We can now use an element
\be
\omega = (\omega_1,\omega_2,\omega_3) \in SU(4) \times SU(2) \times SU(2)
\ee
of order $3$ to break $SU(4) \times SU(2) \times SU(2)$ to
\be
SU(3) \times U(1) \times SU(2) \times U(1).
\ee
Explicitly, we could take
\be
\omega_1 = \text{diag}(\xi,\xi,\xi,1),\qquad
\omega_2 = 1,\qquad
\omega_3 = \text{diag}(\xi,\xi^{-1}),
\ee
with $\xi$ a non-trivial cubic root of unity in $\IC^*$. Combining these two
breaking effects, we conclude that $\alpha \omega$ is an element of order $6$
that breaks $Spin(10)$ to $SU(3) \times SU(2) \times U(1) \times U(1)$.

\subsection{The construction of Standard Model bundles}

We now turn to the essence of this section, which is our attempt at constructing Standard Model bundles on the four $X$'s with $\pi_1(X) = \IZ_6$ constructed above. As usual, we construct a stable bundle $\tV$ on $\tX$, which is invariant under the $\IZ_6$ automorphism group $\bra \t \ket$, so that it descends to a bundle on $X$. As we have seen, to get realistic physics we require $\tV$ to be either an $SU(4)$ or an $SU(5)$ bundle.

The vector bundle $V$ must satisfy the consistency and phenomenological constraints described in section \ref{s:constraints}. For the bundle $\tV$, these translate into the conditions:
\begin{itemize}
\item {\bf (I)}: $\tV$ must be $\bra \tau \ket$-invariant.
\item {\bf (A)}: $\tV$ must satisfy the anomaly cancellation condition which states that $c_2(\tX) - c_2(\tV)$ must be an effective class around which $M5$-branes can wrap (in the weakly coupled regime this must be zero).\footnote{In this section we always consider the hidden bundle to be trivial, and we only allow for $M5$-branes in the spectrum.}
\item {\bf (S)}: $\tV$ must be polystable with respect to a K\"ahler class $\omega$ on $\tX$;
\item {\bf (C1)}: $c_1(\tV) = 0$;
\item {\bf (C3)}: $c_3 (\tV) = \pm 36$.\footnote{This is because $c_3(\tV) = 6 c_3(V)$, since $\tV$ is invariant under the free automorphism $\IZ_6$ on $\tX$.}
\end{itemize}

In the remaining of this section we will try in various ways to construct vector bundles $\tV$ with structure group $SU(4)$ or $SU(5)$ satisfying all of the above conditions. 

\subsection{Pure spectral construction}\label{s:pure}

Perhaps the simplest construction is to use the fact that the simply connected Calabi-Yau threefold $\tX$ is elliptically fibered to construct a stable bundle $\tV$ on $\tX$ directly, using the spectral cover construction of \cite{Donagi:1997pb, Friedman:1997yq}, which we now review briefly.

The Fourier-Mukai transform defines an autoequivalence of the derived category of coherent sheaves $D^b (\tX)$, which goes as follows.\footnote{Physically, the Fourier-Mukai transform can be interpreted as the action on the bundles of $T$-duality along the elliptic fibers of $\tX$.} Let $\tV \to \tX$ be a vector bundle of rank $r$ which is semi-stable and of degree zero on each fiber of $\p': \tX \to B$. Then the Fourier-Mukai transform $FM_{\tX}(\tV)$ is a torsion sheaf on $\tX$ supported on a divisor $e: \S \hookrightarrow \tX$ which is finite and of degree $r$ over $B$. Moreover, $FM_{\tX}(\tV)$ has rank one on $\S$; in fact, if $\S$ is smooth, then $FM_{\tX}(\tV)$ is simply the extension by zero of some line bundle $L \in {\rm Pic}(\S)$.

Conversely, let $e: \Sigma \hookrightarrow \tX$ be a smooth divisor which is finite of degree $r$ over the base $B$ of the elliptic fibration $\pi': \tX \to B$, and let $N \in {\rm Pic}(\S)$ be a line bundle on $\Sigma$; the vector bundle $\tV$ is recovered as the Fourier-Mukai transform $\tV = FM_{\tX} (\S,N)$. The bundle $\tV$ constructed in this way has rank $r$, is semistable and of degree $0$ on each fiber of $\pi'$, and has vertical first Chern class.

Any line bundle $\cl \in {\rm Pic}(\tX)$ has the form $\cl = \p'^*L \otimes \p^* L'$, where $L \in {\rm Pic}(B)$ and $L' \in {\rm Pic}(B')$. Denote by $e: \Sigma \hookrightarrow \tX$ the embedding of the divisor $\S$, and let us first assume that the line bundle $N$ is given by $e_* e^* \cl$ for a global line bundle $\cl \in {\rm Pic}(\tX)$.

Now consistency condition {\bf (I)} requires $\tV$ to be invariant under the $\IZ_6$ automorphism group of $\tX$ generated by $\t$. Since $\tV$ is manufactured through the spectral cover construction, we would like to translate the invariance condition into a condition on the Fourier-Mukai transformed data $(\S, N)$. It was shown in \cite{DOPW2} that $\tV$ is $\bra \t \ket$-invariant if the three following conditions are respected (where $\t_B^*$ denotes the pullback action on the bundles --- we use the notation of the previous sections for the automorphism):
\begin{equation}\label{e:condglob}
\a_{\tX}(\S) = \S,~~~~ \t_B^* L = L,~~~~ {\bf T}_{B'} L' = L' ,
\end{equation}
where $\a_{\tX}$ is the linearization of the automorphism $\t$, which preserves the zero section $\sigma_{B}: B \to \tX$, and ${\bf T}_{B'}$ is the spectral automorphism. By spectral automorphism we mean the action of the automorphism $\t_{B'}$ on the Fourier-Mukai transformed data, which is given by
\begin{equation}
{\bf T}_{B'} = FM_{B'}^{-1} \circ \t_{B'}^* \circ FM_{B'}.
\end{equation}

The two first conditions in \eqref{e:condglob} are relatively easy to satisfy. However, the third condition is subtler. In order to study this condition we need to understand the spectral automorphism ${\bf T}_{B'}$. In fact, we will not need the full spectral automorphism, but rather its cohomological version ${\bf t}_{B'}$ which acts on cohomology classes rather than line bundles. It is defined by
\begin{equation}
{\bf t}_{B'} = fm_{B'}^{-1} \circ \t_{B'}^* \circ fm_{B'},
\end{equation}
where $fm_{B'}$ is the Fourier-Mukai cohomological transform.

In section \ref{s:cohom} we studied the cohomology of the rational elliptic surface $B'$. We found that the cohomological automorphism $\t^*_{B'}$ has a two-dimensional invariant subgroup, generated by the fiber class $[f]$ and the class
\be
[\l] = \sum_{i=0}^5 [e_i],
\ee
where the $e_i$'s are six disjoint sections, and $e_0$ is the zero section. The cohomological Fourier-Mukai transform $fm_{B'}$ was studied in \cite{DOPW} (see table 2, p.32). It is straightforward to show that their result is still valid in our case, where their $e_9$ corresponds to the zero section, which we denoted by $e_0$. Combining this with the cohomological automorphism $\t^*_{B'}$ we obtain the action of ${\bf t}_{B'}$ on cohomology which is shown in table \ref{t:cohact}.

\begin{table}
\centering
\begin{tabular}{|c|c|c|c|c|}
\hline
 & $\t^*_{B'}$ & $fm_{B'} $&$fm_{B'}^{-1}$  &${\bf t}_{B'}$\\
\hline\hline
$1$ & $1$ & $- e_0 + 1/2 \pt$ & $e_0 + 1/2 \pt$& $1 +e_0-e_1+f-\pt$ \\
$\pt$ & $\pt$& $f$&$-f$ & $\pt$ \\
$f$&$f$&$-\pt$ & $\pt $& $f$\\
$e_0$  & $e_1$ & $1-1/2 f$ & $ -1-1/2 f $& $e_0$ \\
$e_1$ & $e_2$ & $1+e_1-e_0-3/2 f- \pt$ & $-1+e_1-e_0-3/2 f + \pt $& $e_2-e_1+e_0+f-\pt$\\
$e_2$ & $e_3$ & $1+e_2-e_0-3/2 f- \pt$ & $-1+e_2-e_0-3/2 f + \pt $&$e_3-e_1+e_0+f-\pt$\\
$e_3$ & $e_4$ & $1+e_3-e_0-3/2 f- \pt$ & $-1+e_3-e_0-3/2 f + \pt $&$e_4-e_1+e_0+f-\pt$\\
$e_4$ & $e_5$ & $1+e_4-e_0-3/2 f- \pt$ & $-1+e_4-e_0-3/2 f + \pt $&$e_5-e_1+e_0+f-\pt$\\
$e_5$ & $e_0$ & $1+e_5-e_0-3/2 f- \pt$ & $-1+e_5-e_0-3/2 f + \pt $&$2 e_0-e_1+2f-2\pt$\\
\hline
$\l$ & $\l$ & $6 + \l - 6 e_0 - 8 f - 5 \pt$& $-6 + \l - 6 e_0 - 8 f + 5 \pt$& $\l + 6 e_0 - 6 e_1 + 6 f - 6 \pt $ \\
\hline
\end{tabular}
\caption{The action of ${\bf t}_{B'}$ on cohomology.}
\label{t:cohact}
\end{table}

Now the invariance condition ${\bf T}_{B'} L' = L'$ implies the cohomological version ${\bf t}_{B'} ({\rm ch}(L') ) = {\rm ch}(L')$, where ${\rm ch}(L') = 1 + c_1(L') + \frac{1}{2} c_1(L')^2$ is the Chern character of the line bundle $L'$. Let $\ell = c_1(L')$. Then $\ell^2$ is a multiple of a point, so is ${\bf t}_{B'}$ invariant; so we need that $1+\ell$ be ${\bf t}_{B'}$ invariant. Since $[\lambda]-6$ is ${\bf t}_{B'}$ invariant by table \ref{t:cohact}, the condition is equivalent to $\ell+ \frac{1}{6}[\lambda]$ being ${\bf t}_{B'}$ invariant. 

Now a class $[x]$ is ${\bf t}_{B'}$ invariant if and only if $fm_{B'} (x)$ is $\t^*_{B'}$ invariant. We know that over the rationals, the
$\t^*_{B'}$ invariant classes (in the full 12-dimensional $H^*(B)$) are spanned by:
\be
1, \pt, [f], [\lambda]. 
\ee
It follows from table \ref{t:cohact} that $[x] = [\ell] + \frac{1}{6}[\l]$ is ${\bf t}_{B'}$ invariant if and only if $fm_{B'}(x)$ is $\t^*_{B'}$ invariant, which is true if and only if $fm_{B'}(x)$ is in the rational span of 
\be
\langle 1, \pt, [f], [\l] \rangle.
\ee 
But using table \ref{t:cohact} again, we see that this is true if and only if $[x] = [\ell] + \frac{1}{6}[\l]$ is in the rational span of
\be
\langle [e_0], \pt, [f], [\l]-6 \rangle.
\ee
In turn, this will true if and only if $\ell$ is in the affine subspace
\be
- \frac{1}{6}[\l] + \text{rational span of } \langle [e_0], [f] \rangle.
\ee
But this does not intersect ${\rm Pic}(B') \subset {\rm Pic}(B') \otimes \IQ$, since $\frac{1}{6} [\l]$ is not an integral class. In other words, what we just showed is that there is no ${\bf T}_{B'}$-invariant line bundles $L' \in {\rm Pic}(B')$.

This implies that the vector bundle $\tV$ cannot be $\bra \t \ket$-invariant if $N = e_* e^* \cl$ for a global line bundle $\cl \in {\rm Pic}(\tX)$. But as explained in \cite{DOPW2}, for a smooth and very ample divisor $\S$ the Lefschetz hyperplane theorem says that every $N \in {\rm Pic}(\S)$ comes from a global line bundle $\cl$, and so we are forced to work with singular or not very ample spectral surfaces $\S$.

Instead of doing so, we will follow the route paved in \cite{DOPW,DOPW2} and build $\tV$ as an extension of two vector bundles $V_1$ and $V_2$, manufactured through the spectral construction. 

\subsection{Spectral construction and extensions}

We define the vector bundle $\tV$ by the short exact sequence
\begin{equation}\label{e:ext}
0 \rightarrow V_1 \rightarrow \tV \rightarrow V_2 \rightarrow 0,
\end{equation}
where the bundles $V_i$, $i=1,2$ are defined by
\begin{equation}\label{e:formV}
V_i = \p'^* W_i \otimes \p^* L_i;
\end{equation}
the $L_i$ are some line bundles on $B'$ and the $W_i$ are bundles on $B$. We will take $W_1$ to be of rank $2$, and $W_2$ to be of rank $r=2$ or $3$, such that $\tV$ has rank $4$ or $5$ respectively. In the latter case, note that we do not lose generality by choosing $W_1$ to be of rank $2$ and $W_2$ of rank $3$, since inverting this choice only means considering the dual bundle.

The $W_i$ are given by the Fourier-Mukai transforms $W_i = FM_B (C_i,N_i)$, where the $C_1, C_2 \subset B$ are smooth curves of degree $2$ and $r$ respectively over the base $\IP^1$ of the elliptic fibration $\beta: B \to \IP^1$, and the $N_i \in {\rm Pic}(C_i)$ are line bundles over $C_i$. 

\subsubsection{Invariance}\label{s:invar}

The first consistency requirement {\bf (I)} is that the bundle $\tV$ must be invariant under the $\IZ_6$ automorphism group generated by $\t: \tX \to \tX$. For $\tV$ to be invariant it is necessary that $V_1$ and $V_2$ also be $\bra \t \ket$-invariant. However, this is not sufficient; we must also ensure that the space $H^1(\tX, V_2^* \otimes V_1)$ parameterizing extensions has a non-zero $\bra \t \ket$-invariant subspace. But for now let us focus on the invariance of $V_1$ and $V_2$.

By construction \eqref{e:formV}, we have that
\begin{equation}
\t^*(V_i) = \p'^* \t_B^* W_i \otimes \p^* \t_{B'}^* L_i,
\end{equation}
for $i=1,2$. Thus it suffices to have a $\bra \t_B \ket$-invariant $W_i$ and a $\bra \t_{B'} \ket$-invariant $L_i$.

For the line bundles $L_i$, we showed in section \ref{s:cohom} that there is a two-dimensional subspace of $H^2(B',\IZ)$ which is $\bra \t_{B'} \ket$-invariant, generated by the classes $[f']$ and $[\l']$ -- recall that the invariant class $[\l']$ is given by the sum of the classes of six disjoint sections, 
\be
[\l'] = \sum_{i=0}^5 [e_i'].
\ee
Thus, if we choose the line bundles to be of the form
\begin{equation}
L_i = \co(a_i [\l'] + b_i [f']),
\end{equation}
for some integers $a_i$ and $b_i$, they will be $\bra \t_{B'} \ket$-invariant as required. In fact, the $[f']$ part can be reabsorbed in the definition of the  bundles $W_i$, and without loss of generality the $\bra \t_{B'} \ket$-invariance of the line bundles implies that
\begin{equation}
L_i = \co (a_i [\l'])
\end{equation}
for some integers $a_i$.

To analyze the $\bra \t_B \ket$-invariance of the higher rank bundles $W_i$ we must work out the action of the automorphism $\t_B$ on the Fourier-Mukai transformed data, as in the previous section. Define the spectral automorphism ${\bf T}_B$ by
\begin{equation}
{\bf T}_B = FM_B^{-1} \circ \t_B^* \circ FM_B.
\end{equation}
From \cite{DOPW2} we know that, for any curve $C \subset B$ which is finite over $\IP^1$,  ${\bf T}_B$ induces a well-defined map ${\bf T}_C : {\rm Pic}(C) \to {\rm Pic}(\a_B C)$, where $\a_B$ is the automorphism of $B$ which preserves the zero section as before. It was shown in \cite{DOPW,DOPW2} that $\bra \t_B \ket$-invariance of $W_i$ reduces to the two following conditions on the spectral data $(C_i,N_i)$:
\begin{equation}
\a_B (C_i) = C_i,~~~~~~~ {\bf T}_{C_i} (N_i) = N_i .
\end{equation}
Solving these constraints is rather involved, and often requires long calculations in the derived category, as in \cite{DOPW,DOPW2}. 
In principle, one first needs to solve the combinatorial problem of finding the right cohomology classes. Given classes that provide a solution at the cohomological level, we still need to find within those classes appropriate invariant curves $C_i$ and invariant line bundles $N_i \in {\rm Pic}^{d_i}(C_i)$ of a specified degree $d_i$.
In the situation explored in this paper we will see that there is a divisibility obstruction already to the initial problem of finding the appropriate cohomology classes.

\subsubsection{Digression: Effective curves}

Before we move on, let us review a few facts about effective curves and their $\a_B$-invariance. First, a divisor $C \subset B$ is effective if and only if $H^0 (B, \co(C) ) \neq 0$. Let $C$ be an effective curve on $B$, and $D$ be an irreducible effective curve on $B$. On the one hand, if $D$ is not a component of $C$; then the intersection number $C \cdot D \geq 0$, since they must intersect in a certain number of points. On the other hand, if $C \cdot D <0$, then $D$ must be a component of $C$; in other words, in this case the curve $C$ is reducible and $C-D$ must also be effective. These simple observations will be useful in the following.

Let the curves $C_i$ lie in the linear systems $LS_i$. By definition, the linear systems $LS_i$ must be effective. Moreover, for $\a_B$-invariant curves $C_i$ to exist, a necessary condition is that the linear systems $LS_i$ must be $\a_B$-invariant themselves. This condition is also sufficient, since the linear systems are projective spaces and any automorphism of a projective space must have fixed points. So $\alpha_B$-invariant curves $C_i$ in $LS_i$ exist if and only if the linear systems $LS_i$ are $\alpha_B$-invariant.
 
In the case under study, the simplest $\a_B$-invariant linear systems that we can consider are the following:
\be
LS_1 = |2 e_0 + k_1 f|, \qquad LS_2 = |r e_0 + k_2 f|, \qquad r=2,3, \quad k_2,k_3 \in \IZ.
\ee
The coefficients of $e_0$ must be $2$ and $r$ so that the curves $C_i$ are $2$-fold and $r$-fold covers of the base $\IP^1$, as required. Moreover, since $LS_i$ must be effective, we have $k_i \geq 0$. In fact, we also assume that the curves $C_i$ are smooth, which implies, using Bertini's theorem, that $k_1 \geq 2$ and $k_2 \geq r$ \cite{DOPW2}.\footnote{Perhaps we should be a little more precise here. Note that from Bertini's theorem we know that for every $k_1 \geq 2$, $k_2 \geq r$ the generic curves in the linear systems $LS_i$ are smooth. However, given a $k$, it is possible that all smooth curves in the linear system $LS$ are not $\a_B$-invariant, and that the only $\a_B$-invariant curves in $LS$ are not smooth. An example of this situation was studied in detail in \cite{DOPW2}, where an $\a_B$-invariant reducible curve which is not finite over the base was considered. But this will not bother us in the following, as all the statements we will make are cohomological. Thus, it is sufficient for us that our $\a_B$-invariant curve $C$ can be deformed to a smooth curve (but not necessarily $\a_B$-invariant) in the same linear system $LS$. 
}

\subsubsection{Chern classes}

Let $C_1 \in |2 e_0 + k_1 f|$, for $k_1 \geq 2$, $C_2 \in |r e_0 + k_2 f|$, with $r=2,3$ and $k_2 \geq r$, and $N_i \in {\rm Pic}^{d_i} (C_i)$, where $d_i$ is the degree of the line bundle. Also, let $L_i = \co (a_i [\l'])$. 

It is straightforward to compute the Chern character of $\tV$ from the results of \cite{DOPW2}. Using the intersection numbers \eqref{e:intfull}, we obtain
\begin{align}
ch(\tV) &= (2 + r) + (2a_1+r a_2) [\pi^* \l']+ \left(d_1 -2 k_1 + 1+d_2 - r k_2 + \frac{1}{2} r(r-1) \right) [\pi^* f'] \notag\\
& \quad +  6\left(- \left(a_1^2+ \frac{r}{2}a_2^2 \right) + a_1(d_1-2k_1+1) + a_2\left(d_2- r k_2+\frac{1}{2} r(r-1) \right) \right)[f \times \pt]\notag\\
&\quad -(k_1+k_2)[\pt \times f']  - 6 (a_1 k_1+a_2 k_2) [\pt] .
\end{align}

\subsubsection{Phenomenological constraints}

We first use the phenomenological constraints {\bf (C1)} and {\bf (C3)}, which give three equations, to express the integers $(a_2,k_2,d_2)$ in terms of $(a_1,k_1,d_1)$. The constraint {\bf (C1)} implies that 
\begin{equation}
\label{e:c1}
2 a_1 = - r a_2,~~~~~ d_1 -2 k_1 + 1+d_2 - r k_2 + \frac{1}{2} r(r-1) = 0,
\end{equation}
while the constraint {\bf (C3)} implies that 
\begin{equation}\label{e:c3}
 a_1 k_1 + a_2 k_2 = \mp 3 .
\end{equation}
Combining these three equations, we get
\begin{align}\label{e:eqnsfor2}
a_2 &= - \frac{2 a_1}{r} , \notag\\
k_2 &= \frac{r}{2} \left( \pm  \frac{3}{a_1} + k_1 \right),\notag\\ 
d_2 &= -d_1+2k_1-1 + \frac{r}{2} \left( r k_1 -r + 1 \pm \frac{3 r}{a_1} \right). 
\end{align}
Thus, in the following we only work with the integers $(a_1,k_1,d_1)$, understanding that $(a_2,k_2,d_2)$ are fixed by \eqref{e:eqnsfor2} (up to the $\pm$ sign, which is fixed by the requirement that $k_2 \geq r$).

Note that integrality of $a_2$, $k_2$ and $d_2$ implies the integrality constraints
\begin{gather}
\frac{2 a_1}{r} \in \IZ,\notag\\
\frac{r}{2} \left( \pm  \frac{3}{a_1} + k_1 \right) \in \IZ.
\end{gather}

\subsubsection{Anomaly cancellation}

The remaining consistency conditions provide inequalities on the integers $(a_1,k_1,d_1)$. Let us first analyze the anomaly constraint {\bf (A)}.

Using the fact that 
\be
c_2(\tX) = 12[f \times \pt]+12[\pt \times f'],
\ee 
we obtain that {\bf (A)} requires the two inequalities
\begin{gather}
k_1 + k_2 \leq 12,\notag\\ \label{e:anomaly}
- \left(a_1^2+ \frac{r}{2}a_2^2 \right) + a_1(d_1-2k_1+1) + a_2\left(d_2- r k_2+\frac{1}{2} r(r-1) \right) \geq -2 .
\end{gather}
Using the phenomenological constraints \eqref{e:eqnsfor2}, these inequalities become
\begin{gather}\label{e:anomaly1}
k_1 \leq \frac{2}{2+r} \left( 12 \mp \frac{3 r}{2 a_1} \right),\notag\\
d_1 - 2 k_1 + 1 \geq - \frac{2 r}{ a_1 (r+2)} + a_1 .
\end{gather}

\subsubsection{Stability}

We now study the stability condition {\bf (S)}. The slope $\mu_{\o}(\tV)$ of a vector bundle $\tV$ on $\tX$ with respect to a K\"ahler class $[\omega] \in H^2(\tX,\IZ)$ is given by
\begin{equation}
\m_{\o}(\tV) = \frac{c_1(\tV) \cdot [\o]^2}{{\rm rk}(\tV)}.
\end{equation}
In particular, if $c_1(\tV) =0 $ then $\m_{\o}(\tV)=0$ for any $\o$. As explained in section \ref{s:constraints}, a vector bundle $\tV$ is stable with respect to $[\o]$ if for all sub-bundles $W$ of $\tV$ we have $\m_{\o}(W) < \m_{\o}(\tV)$.

Recall from \cite{Friedman:1997yq} that starting with any polarization $[\omega_0] \in H^2(\tX, \IZ)$, we can construct a polarization $[\omega]$
\begin{equation}\label{e:polar}
[\omega] = [\omega_0] + m [\pi^* h']
\end{equation}
for some fixed polarization $[h'] \in H^2(B',\IZ)$ and $m \gg 0$ such that every vector bundle $\tV$ on $\tX$ constructed from an irreducible spectral cover is stable with respect to $[\omega]$. 

For the bundle $\tV$ constructed as in \eqref{e:ext}, with the bundles $V_i$ of the form \eqref{e:formV}, it was shown in \cite{DOPW2} that the bundle $\tV$ is stable with respect to the polarization $[\o]$ in \eqref{e:polar} if and only if the extension \eqref{e:ext} is non-split, that is $H^1(\tX, V_2^* \otimes V_1) \neq 0$, and $\m_{\o} (V_1) < 0$. In fact, following the arguments of Lemma 5.4 of \cite{DOPW2}, we can show that the first condition on the extension space is equivalent to $L_1 \cdot f' > L_2 \cdot f'$, that is, $a_1 > a_2$. Using \eqref{e:eqnsfor2}, this implies that
\be
a_1 > 0.
\ee

The second condition, $\m_{\o} (V_1) < 0$, may be restated in a simpler form. We computed above that
\begin{equation}
c_1(V_1) = 2a_1 [\pi^* \l']+ (d_1-2k_1+1) [\pi^* f'].
\end{equation}
Following the arguments in section 5.2 of \cite{DOPW2}, the condition $\m_{\o} (V_1) < 0$ will be satisfied provided that
\begin{equation}\label{e:ample}
(2a_1 [\l']+ (d_1-2k_1+1) [f']) \cdot [h'] < 0
\end{equation}
for some ample class $[h'] \in H^2(B',\IZ)$. Now, since the K\"ahler cone is dual to the Mori cone it is clear that the existence of an ample class $[h']$ satisfying \eqref{e:ample} is equivalent to the statement that the class
\begin{equation}\label{e:effec}
[\n] = 2a_1 [\l']+ (d_1-2k_1+1) [f']
\end{equation}
is not in the Mori cone of $B'$, that is, it is not effective. 

How do we know when the class $[\n]$ is effective or not? Recall that $[\l']$ is the sum of the classes of six disjoint sections $[e_i']$, $i=0,\ldots,5$. Intersecting $[\n]$ with the class of the zero section $[e_0']$, we get
\be
[\n] \cdot [e_0'] = - 2 a_1 + (d_1-2k_1+1).
\ee
Assume that $[\n]$ is effective. Let
\be
d_1-2k_1+1= - b < 0.
\ee
Then, 
\be
[\n] \cdot [e_0'] = - b -2 a_1< 0
\ee
since $a_1>0$,
and so 
\be
[\n_1]:= [\n] - (b+2 a_1) [e_0']
\ee
must be effective. Intersect now $[\n_1]$ with $[e_1']$:
\be
[\n_1] \cdot [e_1']= - (b+2 a_1) <0,
\ee
since $[e_0']\cdot [e_1'] = 0$. Hence
\be
[\n_2]:=[\n] - (b+2 a_1) \left([e_0']+[e_1']\right)
\ee
must also be effective. Iterating with the other $e_i's$ up to $e_5'$, we end up with the condition that
\begin{align}\label{e:n6}
[\n_6]:=&[\n] - (b+2a_1)[\l']\notag\\
 =& - b[\l'] - b [f']
\end{align}
must also be effective, which is clearly wrong, since $b>0$.
Therefore, we get that for
\be
d_1 - 2 k_1 \leq -2 ,
\ee
the class $[\n]$ is not effective.

To summarize, consistency conditions of the compactification require the four following inequalities on $(a_1,k_1,d_1)$.

\paragraph{Anomaly cancellation:}
\begin{gather}
k_1 \leq \frac{2}{2+r} \left( 12 \mp \frac{3 r}{2 a_1} \right),\notag\\\label{e:anomfin}
d_1 - 2 k_1 + 1 \geq - \frac{2 r}{ a_1 (r+2)} + a_1 .
\end{gather}

\paragraph{Stability:}
\begin{gather}
a_1 > 0, \notag\\
d_1 - 2 k_1 \leq -2 .\label{e:stabfin}
\end{gather}

Moreover, we have by definition that $k_1 \geq 2$, and $r=2$ or $3$. We also have that $k_2 \geq r$, which implies, through \eqref{e:eqnsfor2} that
\be\label{e:ineqk2}
 k_1 \geq 2  \mp  \frac{3}{a_1} .
\ee
There is also the integrality constraints:
\begin{gather}
\frac{2 a_1}{r} \in \IZ,\notag\\\label{e:integfin}
\frac{r}{2} \left( \pm  \frac{3 }{a_1} + k_1 \right) \in \IZ.
\end{gather}
For $r=2$, the first one is trivially satisfied for $a_1 \in \IZ$, while the second one implies that $a_1$ must divide $3$. For $r=3$, the first constraint imposes that $a_1$ be a multiple of $3$, while the second constraint imposes that $a_1$ divide $9$, and that $\frac{9 }{a_1} + 3k_1$ be even.

\subsubsection{Solutions?}

To study the inequalities, we proceed as follows. We consider $r=2$ and $3$ independently. For each of these, we first use the integrality constraints \eqref{e:integfin} and the first stability condition \eqref{e:stabfin} to find the allowed values for $a_1$. Then, we use the second stability condition \eqref{e:stabfin} and the second anomaly condition \eqref{e:anomfin} to find allowed values of $d_1$ and $k_1$:
\be
\label{e:ineqhard}
- \frac{2 r}{ a_1 (r+2)} + a_1 \leq d_1 - 2 k_1 + 1 \leq -1.
\ee
This inequality is very interesting, since it shows explicitly the clear tension between the anomaly cancellation condition and the stability condition. It turns out to be impossible to find $d_1$'s and $k_1$'s satisfying these two inequalities.

\begin{list}{$\bullet$}{\setlength{\leftmargin}{0.4cm}}
\item
\boxed{r=2} The allowed values of $a_1$ are $1$ or $3$. However, inequality \eqref{e:ineqhard} is never satisfied, since the LHS is strictly greater than the RHS. Indeed, for $r=2$, the LHS is
\be
- \frac{1}{a_1} + a_1,
\ee
which is for both $a_1=1$ and $3$ strictly greater than the RHS, $-1$. Therefore, there are no solutions to the stability and anomaly cancellation constraints for $r=2$, and we cannot construct globally consistent $SU(4)$ bundles.

Notice however the strong tension between anomaly cancellation and stability. If we drop the anomaly cancellation condition, arguing as in section 4, the left inequality of \eqref{e:ineqhard} disappears, and we get an infinite family of solutions, as in the $\IZ_2$ construction. However, in contrast with the $\IZ_2$ construction, here there is no unique solution satisfying both global consistency conditions at the same time.

\item
\boxed{r=3} Here, $a_1 = 3$ or $9$. But again \eqref{e:ineqhard} is never satisfied, since the LHS
\be
 - \frac{6}{5 a_1} + a_1
\ee
is strictly greater than $-1$ for both $a_1=3$ and $9$. So we cannot construct globally consistent $SU(5)$ bundles either. Notice again however that if we drop anomaly cancellation, we get an infinite family of phenomenologically consistent vacua.

\end{list}

\subsection{Conclusions}

Let us conclude this section with a few comments. We tried to solve the numerical constraints coming from anomaly cancellation, stability of the vector bundle, and the phenomenological requirements. We found no solutions. However, we obtained a result similar to what we discussed in the previous section. Namely, if we drop the topological anomaly condition, for instance by allowing anti-branes, or by focusing locally on the visible brane, we get plenty of models, infinite families of models in fact, phenomelogically viable at the topological level. But if we consider globally consistent compactifications, either perturbative or at strong coupling with $M5$-branes, we get no solutions. In other words, none of these local models can be UV completed into globally consistent supersymmetric string vacua. 

However, there are various ways to generalize the bundle construction of the previous section to try to get Standard Models on these $\IZ_6$ threefolds. One possibility would be to consider spectral curves $C_i$ lying in linear systems $LS_i$ larger than $|2 e_0 + k_1 f|$ and $|r e_0 + k_2 f|$. For instance, we could consider linear systems $LS_i$ given by $LS_i=|r_i e_0 +k_i f + \sum_n a_{i,n} \m_n|$, where the $\m_n$ are some $\a_B$-invariant classes. We tried in various ways to construct bundles using these extended linear systems, without success. We also tried to consider reducible curves $C_i$, but ran into problems. 

This is not to say that we are ready to rule out the construction of Standard Model on these $\IZ_6$ threefolds. Rather, we state that the type of construction that has been used in the models of \cite{DOPW,DOPW2,BD} does not seem to yield viable bundles in this case; and we have not suceeded yet in generalizing it to produce suitable bundles. It would of course be more satisfactory to produce a systematic study of the moduli space of bundles on these threefolds, in order to prove the existence or non-existence of Standard Model bundles on these manifolds.

It would also be desirable to study the construction of bundles on the other non-simply connected Calabi-Yau threefolds in the classification of \cite{BD2}; perhaps some of these threefolds will yield realistic vacua. Indeed, while only the threefolds with $\IZ_6$ and $\IZ_3 \times \IZ_3$ fundamental groups can be used to break the $Spin(10)$ gauge group to the MSSM gauge group with an extra $U(1)_{B-L}$, all the threefolds can be used to break $SU(5)$ to the MSSM gauge group. It is conceivable that Standard Model bundles similar to the one constructed in \cite{BD} exist on other threefolds.

\appendix

\section{Computation of the invariant cohomology group $H^2(B,\IZ)^{\langle \t_B \rangle}$}

In this appendix, we compute the invariant cohomology group $H^2(B,\IZ)^{\langle \t_B \rangle}$ for the four families of rational elliptic surfaces that are used to construct the four non-simply connected Calabi-Yau threefolds $X$ with fundamental group $\pi_1(X)=\IZ_6$. 
These invariant classes were needed in section 5.2.
We first outline a general procedure for computing the invariant cohomology group for any rational elliptic surface with a finite group of automorphisms classified in \cite{BD2}. We then apply our procedure to the four cases under consideration.

\subsection{Outline of the procedure}

Let us recall the details of the geometry. We are given a non-generic rational elliptic surface $B$, with an automorphism $\tau_B : B \to B$ of order $n$ (take any case in tables 6, 8 and 9 of \cite{BD2}). The automorphism $\tau_B$ has the form
\be
\tau_B = t_\xi \circ \alpha_B,
\ee
where $\alpha_B : B \to B$ is the linearization of $\tau_B$ which fixes the zero section $\sigma : \IP^1 \to B$ of the elliptic fibration of $B$, and $t_\xi$ denotes translation by a section $\xi$ in the Mordell-Weil group $MW(B)$. We denote the order of $\alpha_B$ by $m$, with $m | n$.

Let $\t_{\IP^1}: \IP^1 \to \IP^1$ be the automorphism induced by $\t_B$. When it is non-trivial, denote by $0,\infty \in \IP^1$ its two fixed points. By construction, we know that $\alpha_B$ fixes the fiber over $\infty$, which we denote by $f_\infty$, pointwise, and has isolated fixed points in $f_0$.

Using the notation of \cite{BD2}, we define the group homomorphism $\cp_i:MW(B) \to MW(B)$:
\be
\cp_i: \xi \to \cp_i(\xi):= \sum_{j=0}^{i-1} \alpha_B^{j} \xi.
\ee
For $\t_B$ to be of order $n$, the section $\xi$ must satisfy
\be
\cp_n(\xi) = \sigma,
\ee
that is, it must be in the kernel of $\cp_n$. We must also require that the section $\xi$ intersects $f_\infty$ at a torsion point of order precisely $n$, for $\langle \t_B \rangle$ to act freely on $f_\infty$.

Not all choices of $\xi$'s however lead to inequivalent automorphisms. In particular, we proved in \cite{BD2} that all sections in a given coset of $\Im(1-\alpha_B)$ in $\ker(\cp_n)$ lead to conjugate automorphisms. This means that we do not have to consider all $\xi$'s in $\ker (\cp_n)$, but can restrict ourselves to nice representatives of the cosets. Moreover, we showed that all sections in a given coset intersect $f_\infty$ at the same torsion point (or 0). Hence we may only consider representatives of the cosets intersecting $f_\infty$ at a torsion point of order precisely $n$.

We want to compute the induced automorphism $\t_B^*$ on the cohomology group $H^2(B, \IZ)$, and its invariant subgroup. Define the subgroup $T \subset H^2(B, \IZ)$ generated by the fiber class $[f]$, the class of the zero section $[\sigma]$, and the classes of the non-neutral components of the singular fibers of $B$. There is a well known short exact sequence
\be
0 \to T \to H^2(B,\IZ) \to MW(B) \to 0,
\ee
which tells us that $H^2(B,\IZ)$ is generated by the fiber class, the class of $\sigma$, the classes of the non-neutral components of the singular fibers, and the classes of sections generating $MW(B)$.

The action of $\t_B^*$ on the classes generating $T$ is straightforward to compute. We know that $\alpha_B$ fixes $[f]$ and $[\sigma]$, and its action on the singular fibers follows from the geometry of $B$. Then, $\tau_B^*$ fixes $[f]$, sends $[\sigma]$ to $[\xi]$, and we can compute its action on the classes of the non-neutral components using the intersection numbers of $\xi$.

What remains to be done is the computation of $\t_B^*$ on the classes of sections generating $MW(B)$. The action of translation by a section $\xi$ is clear; what we need to compute is the action of $\alpha_B$, when it is non-trivial. Since $\alpha_B$ fixes the zero section, it defines an automorphism of the Mordell-Weil group $MW(B)$, which projects onto an automorphism of the Mordell-Weil lattice $MW_{lat}(B)$. Moreover, in all cases in tables 6, 8 and 9 of \cite{BD2}, we know that $\alpha_B$ fixes the torsion subgroup of the Mordell-Weil group. We also know the fixed locus of $\alpha_B$ on $B$; it fixes the smooth elliptic fiber $f_\infty$ pointwise, and has a certain number of isolated fixed points on $f_0$. We can then use the Lefschetz fixed-point theorem to compute the trace of $\alpha_B$ on $H^2(B,\IQ)$; using the explicit action on $T$, we extract the trace of $\alpha_B$ on the Mordell-Weil lattice $MW_{lat}(B)$.

From this information we would like to write down an explicit matrix for $\alpha_B$ on a given basis of $MW_{lat}(B)$.
Let ${\rm Aut}(MW_{lat})$ be the automorphism group of the Mordell-Weil lattice. Generally, it will be given by the extension of the Weyl group $W(MW_{lat})$ by the outer automorphisms of the Dynkin diagram. Conjugacy classes in Weyl groups have been classified; see for instance \cite{Carter, Dol}. Moreover, automorphisms of Dynkin diagrams are relatively simple. Hence, using what we know about $\alpha_B$, namely its order, its trace and its invariant sublattice, we should be able to identify in which conjugacy class in ${\rm Aut}(MW_{lat})$ it lies. We can then write down an explicit matrix representative in this class. 

Using this explicit representative for $\alpha_B$, we can write down the map from $\ker(\cp_n)$ to the quotient group 
\be
\frac{\ker(\cp_n)}{\Im(1-\alpha_B)}.
\ee 
We determine which cosets intersect $f_\infty$ at a torsion point of order $n$. Using representatives for these cosets, we can work out in detail the invariant cohomology $H^2(B,\IZ)^{\langle \t_B \rangle}$ for all possible choices of suitable $\xi$'s.

\subsection{The $\IZ_6$ cases}

Let us now exemplify this procedure for the four cases with order 6 automorphisms $\tau_B$, corresponding to case 3 of table 6 of \cite{BD2}, and cases 13, 14 and 15 of table 8.

Recall that in these four cases, we know that
\be
\dim H^2(B,\IZ)^{\langle \t_B \rangle} = 2.
\ee
We also constructed two invariant classes of $H^2(B,\IZ)$:
\be
[f], \qquad [\l] = \sum_{i=0}^5 [e_i],
\ee
where the sections $e_i$ form the $\t_B$-orbit of the zero section $\sigma := e_0$. We know the intersection numbers
\be
[f] \cdot [f] = 0, \qquad [f] \cdot [\l] = 6.
\ee
We need to compute the self-intersection of $[\l]$, and determine whether $[f]$ and $[\l]$ generate the full invariant cohomology, or only a finite index subgroup thereof.

\subsubsection{First family}

We start with case 3 of table 6. $B$ is a rational elliptic surface with configuration of singular fibers $\{ I_6, I_3, I_2, I_1 \}$. The Mordell-Weil group $MW(B)$ has rank $0$, and is given by $MW(B) = \IZ_6$. In other words, the only sections are the zero section $\sigma$, and a torsion section $\eta$ of order $6$, with its multiples. 

For this family, the order 6 automorphism $\t_B : B \to B$ is simply given by translation by the order 6 torsion section $\eta$; that is, the linearization $\alpha_B$ of $\t_B$ is the identity, as is the projection of $\t_B$ to $\IP^1$. Consequently, the invariant class $[\l]$ reads
\be
[\l] = [\sigma] + [\eta] + [2 \eta] + [3 \eta] + [4 \eta]+ [5 \eta].
\ee
Since torsion sections are necessarily disjoint from the zero section, and disjoint from each other, we obtain directly the intersection number
\be
[\l] \cdot [\l] = -6.
\ee

Now we need to figure out whether $[f]$ and $[\l]$ generate the full $H^2(B, \IZ)^{\langle \t_B \rangle}$. In other words, we must determine whether there exists a class
\be
[\r] = a [\l] + b [f],
\ee
with $a, b \in \IZ$ relatively prime, which is divisible. 

Suppose that
\be
\frac{1}{m} [\r]
\ee
is integral for some integer $m \in \IZ$. Intersect with the class of the zero section $[\sigma]$. One gets that
\be\label{e:zz}
\frac{b-a}{m} \in \IZ,
\ee
which implies that
\be
[\r'] = \frac{a}{m} ([f] + [\l] )
\ee
must also be integral. Squaring it we get that
\be
6 \frac{a^2}{m^2} \in \IZ,
\ee
which implies that $m$ must divide $a$, which in turn implies that $m$ must also divide $b$ by \eqref{e:zz}, hence a contradiction. Therefore we get that $[\r]$ is never divisible, and $[f]$ and $[\l]$ generate the full invariant cohomology group $H^2(B, \IZ)^{\langle \t_B \rangle}$. 

Note that this last result does not depend on the particular geometry of the rational elliptic surface, but only on the fact that the class $[\l]$ satisfies $[\l] \cdot [\l] = -6$. Hence for the other families, we will only need to show that $[\l] \cdot [\l] = -6$.

\subsubsection{Second family}

We consider case 15 in table 8. The configuration of singular fibers is $\{ IV, 2I_3, 2I_1 \}$, and the Mordell-Weil group is $MW(B) = A_2^* \oplus \IZ_3$. The order 6 automorphism has the form $\t_B = t_\xi \circ \alpha_B$, where the linearization $\alpha_B$ has order $2$, and the section $\xi \in MW(B)$ satisfies
\be
\label{e:condeta}
\alpha_B \xi + \xi = \eta,
\ee 
where $\eta$ is a 3-torsion section. That is, $\eta \in \ker(\cp_6)$, and intersects $f_\infty$ at a torsion point of order 6. For this family, $\ker(\cp_6) = MW$.

The invariant class $[\l]$ reads
\be
[\l] = [\sigma] + [\xi] + [\eta] + [\xi + \eta] + [2 \eta] + [\xi + 2 \eta].
\ee
$\alpha_B$ has 4 isolated fixed points on the singular fiber of type $IV$ over $0 \in \IP^1$, and fixes the smooth fiber over $\infty \in \IP^1$ pointwise.

We know that $\alpha_B$ fixes the torsion sections, and its action on $MW(B)$ projects onto an automorphism of order 2 of the lattice $A_2^*$. Using Lefschetz fixed-point theorem we get that the trace of $\alpha_B$ on $A_2^*$ is $-2$. We also know that it has no invariant sublattice. Since $A_2^*$ is equivalent to $A_2$ (up to scaling), its automorphism group $\Aut(A_2^*)$ is equal to $\Aut(A_2)$. It is given by the extension of the Weyl group $W(A_2)$ by the automorphism group of the Dynkin diagram, which is just the order 2 group generated by negation of all coordinates in $A_2^*$; let us denote this element by $-1$. Using the classification of conjugacy classes in Weyl groups, we get that $\alpha_B$ is in the conjugacy class containing the element $-1$, which has trace $-2$ and no invariant sublattice. Hence we can take $\alpha_B$ to act as $-1$ on the lattice $A_2^*$.

Using this description of $\alpha_B$ we compute the quotient group
\be
\frac{MW}{\Im(1-\alpha_B)} \simeq (\IZ_2)^2 \times \IZ_3,
\ee 
which has cardinality 12. We also compute explicitly the map $MW \to (\IZ_2)^2 \times \IZ_3$, and show that each non-trivial coset contains at least one representative which is disjoint from the zero section (either a torsion section, or a section which projects to a minimal point of the lattice $A_2^*$, which means, as was shown in \cite{OS, Sh}, that it is disjoint from $\sigma$). This implies that the map from the quotient group $(\IZ_2)^2 \times \IZ_3$ to the torsion points of $f_\infty$ is injective, and we know exactly which cosets intersect $f_\infty$ at torsion points of order precisely 6. 

Each of these suitable cosets contains a representative which projects to a minimal point of $A_2^*$; we define $\xi$ to be such a section. Using the height pairing in the lattice, it is easy to show that $\xi$ must also be disjoint from the torsion sections $\eta$ and $\eta + \eta$, which are also disjoint from the zero section. We end up with the intersection number
\be
[\l] \cdot [\l] = -6,
\ee
for all suitable $\xi$'s, as for the first family. 

\subsubsection{Third family}

We consider case 14 in table 8. The configuration of singular fibers is $\{ III, 3I_2, 3I_1 \}$, and the Mordell-Weil group is $MW(B) = D_4^* \oplus \IZ_2$. The order 6 automorphism has the form $\t_B = t_\xi \circ \alpha_B$, where the linearization $\alpha_B$ has order $3$, and the section $\xi \in MW(B)$ satisfies
\be
\cp_3(\xi) = \eta,
\ee 
where $\eta$ is a 2-torsion section. Here again, $\ker(\cp_6) = MW$. The invariant class $[\l]$ reads
\be
[\l] = [\sigma] + [\xi] + [\cp_2(\xi)] + [\eta] + [\eta + \xi] + [\eta + \cp_2(\xi)].
\ee
$\alpha_B$ has 3 isolated fixed points on the singular fiber of type $III$ over $0 \in \IP^1$, and fixes the smooth fiber over $\infty \in \IP^1$ pointwise. 

We know that $\alpha_B$ fixes the torsion sections, and its action on $MW(B)$ projects onto an automorphism of order 3 of the lattice $D_4^*$. Using Lefschetz fixed-point theorem we get that the trace of $\alpha_B$ on $D_4^*$ is $-2$. We also know that it has no invariant sublattice. Since $D_4^*$ is equivalent to $D_4$ (up to scaling), $\Aut(D_4^*) = \Aut(D_4)$. The latter is given by the extension of the Weyl group $W(D_4)$ by the outer automorphisms of the Dynkin diagram, which in this case is the symmetric group $S_3$ of order 6 ($D_4$ exhibits so-called triality). In fact, $\Aut(D_4)$ can be identified with the Weyl group $W(F_4)$ of the exceptional group $F_4$.

Conjugacy classes in $W(F_4)$ are listed in table 8, p.48 of \cite{Carter}. There is only one class containing an element of order 3 with trace $-2$ and no invariant sublattice. It is indexed by the Carter graph
\be
A_2 \times \tilde A_2,
\ee
which means that a representative is given by the product of an element expressible as a product of reflections corresponding to long roots, and an element expressible as a product of reflections corresponding to short roots. Details aside, the point is that there is only one conjugacy class in $\Aut(D_4)$ with the required properties, and we can write an explicit matrix representative for $\alpha_B$. It is in fact obtained by composing an element of $W(D_4)$ with an outer automorphism.

Using this description of $\alpha_B$ we compute the quotient group
\be
\frac{MW}{\Im(1-\alpha_B)} \simeq (\IZ_3)^2 \times \IZ_2,
\ee 
which has cardinality 18. We also compute explicitly the map $MW \to (\IZ_3)^2 \times \IZ_2$, and show that each non-trivial coset contains at least one representative which is disjoint from the zero section. This implies that the map from the quotient group $(\IZ_3)^2 \times \IZ_2$ to the torsion points of $f_\infty$ is injective, and we know exactly which cosets intersect $f_\infty$ at torsion points of order precisely 6. 

Again, all suitable cosets contains at least one representative which projects to a minimal point of $D_4^*$, hence is disjoint from $\sigma$. We define $\xi$ to be such a section. Using the explicit matrix for $\alpha_B$, it is easy to show that $\cp_2(\xi) = \alpha_B \xi + \xi$ also projects to a minimal point, thus is disjoint from $\sigma$.\footnote{This can also be obtained as follows.
If $[\xi] \cdot [\sigma] = 0$, by $\IZ_6$-symmetry this implies that $[\xi] \cdot [\cp_2(\xi)] = 0$ and $[\cp_2(\xi)] \cdot [\eta]=0$. Moreover, using the height pairing, one can see that the condition $[\xi] \cdot [\sigma] = 0$ implies that $\xi$ must intersect two non-neutral components of the singular fibers. Considering the action of $\alpha_B$ on these components, we compute that $\cp_2(\xi)$ also intersects two non-neutral components. Finally, pairing $\cp_2(\xi)$ with the torsion section $\eta$ and using the fact that the torsion section $\eta$ is disjoint from $\sigma$, we obtain that $\cp_2(\xi)$ must also be disjoint from the zero section.} Both sections are also disjoint from the torsion section $\eta$, and we obtain again that
\be
[\l] \cdot [\l] = -6,
\ee
for all suitable $\xi$'s.

\subsubsection{Fourth family}

We consider case 13 in table 8. The configuration of singular fibers is $\{ 12 I_1 \}$, and the Mordell-Weil group is $MW(B) = E_8$. The order 6 automorphism has the form $\t_B = t_\xi \circ \alpha_B$, where the linearization $\alpha_B$ has order $6$, and the section $\xi \in MW(B)$ satisfies
\be\label{e:cp6}
\cp_6(\xi) = \sigma.
\ee 
Again, $\ker(\cp_6) = MW$.
The invariant class $[\l]$ reads
\be
[\l] = \sum_{i=0}^5 [\cp_i(\xi)].
\ee
$\alpha_B$ has one fixed point on the smooth fiber over $0 \in \IP^1$, and fixes the smooth fiber over $\infty \in \IP^1$ pointwise.

$\alpha_B$ is an automorphism of order 6 of the lattice $E_8$ with no invariant sublattice. Using Lefschetz fixed-point theorem we get that the trace of $\alpha_B$ on $E_8$ is $-3$. The automorphism group of $E_8$ is just the Weyl group $W(E_8)$. Looking at conjugacy classes in $W(E_8)$ (table 11, pp.54-58 of \cite{Carter}), we see that there is only one conjugacy class with element of order 6, trace -3 and no invariant sublattice; it has Carter graph
\be
A_1 \times A_2 \times A_5.
\ee
A representative is given by the product of Coxeter elements of $W(A_1) \times W(A_2) \times W(A_5)$. Hence we can use this representative for our $\alpha_B$. We can write it in matrix form as follows. First, we use the embedding of $A_1 \oplus A_2 \oplus A_5$ in $E_8$, which is a subgroup of index 6, given by removing the central node of the extended Dynkin diagram of $E_8$, to write down $\alpha_B$ as an $8 \times 8$ matrix in the basis of $A_1 \oplus A_2 \oplus A_5$. We then conjugate it with the change of basis to get a matrix form for $\alpha_B$ in the standard basis of $E_8$. 

Using this description, we compute the quotient group
\be
\frac{MW}{\Im(1-\alpha_B)} \simeq (\IZ_6)^2,
\ee
which has cardinality 36. We write down the map $E_8 \to (\IZ_6)^2$, and show that each coset contains a representative disjoint from $\sigma$. Hence the map from the quotient group $(\IZ_6)^2$ to the torsion points of $f_\infty$ is injective, and we know which cosects intersect $f_\infty$ at torsion points of order 6.

Again, each of these suitable cosets contains at least one representative which is a minimal point of $E_8$, hence a section disjoint from $\sigma$. Define $\xi$ to be such a section. One can then check explicitly that the sections $\cp_i(\xi)$, $i=2,\ldots,5$ are also of minimal length, hence disjoint from $\sigma$. We thus obtain again that
\be
[\l] \cdot [\l] = -6,
\ee
for all suitable $\xi$'s.

To conclude, we have showed that for all four families, and for all suitable automorphisms $\tau_B$, the invariant cohomology group $H^2(B, \IZ)^{\langle \t_B \rangle}$ is always generated by two classes $[f]$ and $[\l]$, with intersection numbers
\be
[f] \cdot [f] = 0, \qquad [f] \cdot [\l] = 6, \qquad [\l] \cdot [\l] = -6.
\ee


\begin{thebibliography}{99}
\bibliographystyle{plain}

\bibitem{AD}
  B.~S.~Acharya and M.~R.~Douglas,
  ``A finite landscape?,''
  arXiv:hep-th/0606212.
  %%CITATION = HEP-TH/0606212;%%

\bibitem{AC}
  B.~Andreas and G.~Curio,
  ``Heterotic models without fivebranes,''
  J.\ Geom.\ Phys.\  {\bf 57}, 2136 (2007)
  [arXiv:hep-th/0611309].
  %%CITATION = JGPHE,57,2136;%%

\bibitem{AHR}
  B.~Andreas and D.~Hernandez Ruiperez,
  ``U(n) vector bundles on Calabi-Yau threefolds for string theory
  compactifications,''
  Adv.\ Theor.\ Math.\ Phys.\  {\bf 9}, 253 (2005)
  [arXiv:hep-th/0410170].
  %%CITATION = HEP-TH 0410170;%%

\bibitem{BHV}
  C.~Beasley, J.~J.~Heckman and C.~Vafa,
  ``GUTs and Exceptional Branes in F-theory - I,''
  arXiv:0802.3391 [hep-th].
  %%CITATION = ARXIV:0802.3391;%%

\bibitem{BMRW}
  R.~Blumenhagen, S.~Moster, R.~Reinbacher and T.~Weigand,
  ``Massless spectra of three generation U(N) heterotic string vacua,''
  JHEP {\bf 0705}, 041 (2007)
  [arXiv:hep-th/0612039].
  %%CITATION = JHEPA,0705,041;%%

\bibitem{BMW}
  R.~Blumenhagen, S.~Moster and T.~Weigand,
  ``Heterotic GUT and standard model vacua from simply connected Calabi-Yau
  manifolds,''
  Nucl.\ Phys.\  B {\bf 751}, 186 (2006)
  [arXiv:hep-th/0603015].
  %%CITATION = NUPHA,B751,186;%%

\bibitem{BCD}
  V.~Bouchard, M.~Cveti\v c and R.~Donagi,
  ``Tri-linear couplings in an heterotic minimal supersymmetric standard
  model,''
  Nucl.\ Phys.\  B {\bf 745}, 62 (2006)
  [arXiv:hep-th/0602096].
  %%CITATION = NUPHA,B745,62;%%

\bibitem{BD}
  V.~Bouchard and R.~Donagi,
  ``An SU(5) heterotic standard model,''
  Phys.\ Lett.\  B {\bf 633}, 783 (2006)
  [arXiv:hep-th/0512149].
  %%CITATION = PHLTA,B633,783;%%

\bibitem{BD2}
  V.~Bouchard and R.~Donagi,
  ``On a class of non-simply connected Calabi-Yau threefolds,'' 
  Comm. Numb. Theor. Phys. {\bf 2}, 1 (2008)
  [arXiv:0704.3096 [math.AG]].
  %%CITATION = PHLTA,B633,783;%%

\bibitem{BBO}
  V.~Braun, E.~I.~Buchbinder and B.~A.~Ovrut,
  ``Dynamical SUSY breaking in heterotic M-theory,''
  Phys.\ Lett.\  B {\bf 639}, 566 (2006)
  [arXiv:hep-th/0606166];\\
  %%CITATION = PHLTA,B639,566;%%
  V.~Braun, E.~I.~Buchbinder and B.~A.~Ovrut,
  ``Towards realizing dynamical SUSY breaking in heterotic model building,''
  JHEP {\bf 0610}, 041 (2006)
  [arXiv:hep-th/0606241].
  %%CITATION = JHEPA,0610,041;%%


\bibitem{BHOP}
  V.~Braun, Y.~H.~He, B.~A.~Ovrut and T.~Pantev,
  ``The exact MSSM spectrum from string theory,''
  JHEP {\bf 0605}, 043 (2006)
  [arXiv:hep-th/0512177].
  %%CITATION = JHEPA,0605,043;%%

\bibitem{BO}
  V.~Braun and B.~A.~Ovrut,
  ``Stabilizing moduli with a positive cosmological constant in heterotic
  M-theory,''
  JHEP {\bf 0607}, 035 (2006)
  [arXiv:hep-th/0603088].
  %%CITATION = JHEPA,0607,035;%%

\bibitem{Bu}
  E.~I.~Buchbinder,
  ``Raising anti de Sitter vacua to de Sitter vacua in heterotic M-theory,''
  Phys.\ Rev.\  D {\bf 70}, 066008 (2004)
  [arXiv:hep-th/0406101].
  %%CITATION = PHRVA,D70,066008;%%

\bibitem{BuO}
  E.~I.~Buchbinder and B.~A.~Ovrut,
  ``Vacuum stability in heterotic M-theory,''
  Phys.\ Rev.\  D {\bf 69}, 086010 (2004)
  [arXiv:hep-th/0310112].
  %%CITATION = PHRVA,D69,086010;%%

\bibitem{BHLR}
  W.~Buchmuller, K.~Hamaguchi, O.~Lebedev, M.~Ratz,
 ``Supersymmetric Standard Model from the Heterotic String,''
arXiv:hep-th/0606187.

\bibitem{Carter}
R.~Carter, ``Conjugacy classes in the Weyl group," in \emph{Seminar on Algebraic Groups and Related Finite Groups}, The Institute for Advanced Study, Princeton, NJ, 1968/69, pp.297--318, Springer, Berlin.

\bibitem{CD}
  G.~Curio and R.~Donagi,
  ``Moduli in $N=1$ heterotic/F-theory duality,''
  arXiv:hep-th/9801057.


\bibitem{CLW}
  M.~Cveti\v c, P.~Langacker and J.~Wang,
  ``Dynamical supersymmetry breaking in standard-like models with  intersecting
  D6-branes,''
  Phys.\ Rev.\  D {\bf 68}, 046002 (2003)
  [arXiv:hep-th/0303208].
  %%CITATION = PHRVA,D68,046002;%%


\bibitem{Dol}
I.~V.~Dolgachev and V.~A.~Iskovskikh, ``Finite subgroups of the plane Cremona group," arXiv:math/0610595.

\bibitem{Donagi:1997pb}
R.~Y. Donagi, ``Principal bundles on elliptic fibrations,"
   arXiv:alg-geom/9702002.

\bibitem{DHOR}
  R.~Donagi, Y.~H.~He, B.~A.~Ovrut and R.~Reinbacher,
  ``The spectra of heterotic standard model vacua,''
  JHEP {\bf 0506}, 070 (2005)
  [arXiv:hep-th/0411156].
  %%CITATION = JHEPA,0506,070;%%

\bibitem{DOPW}
  R.~Donagi, B.~A.~Ovrut, T.~Pantev and D.~Waldram,
  ``Spectral involutions on rational elliptic surfaces,''
  Adv.\ Theor.\ Math.\ Phys.\  {\bf 5}, 499 (2002)
  [arXiv:math.ag/0008011].
  %%CITATION = MATH-AG 0008011;%%

\bibitem{DOPW2}
  R.~Donagi, B.~A.~Ovrut, T.~Pantev and D.~Waldram,
  ``Standard-model bundles,''
  Adv.\ Theor.\ Math.\ Phys.\  {\bf 5}, 563 (2002)
  [arXiv:math.ag/0008010].
  %%CITATION = MATH-AG 0008010;%%

\bibitem{DW}
  R.~Donagi and M.~Wijnholt,
  ``Model Building with F-Theory,''
  arXiv:0802.2969 [hep-th].
  %%CITATION = ARXIV:0802.2969;%%

\bibitem{Friedman:1997yq}
R.~Friedman, J.~{M}organ, and E.~{W}itten, ``Vector bundles and {F} theory,"
   Commun. {M}ath. {P}hys. {\bf 187}, 679--743 (1997)
  [arXiv:hep-th/9701162].

\bibitem{GLS}
  T.~L.~Gomez, S.~Lukic and I.~Sols,
  ``Constraining the Kaehler moduli in the heterotic standard model,''
  Commun.\ Math.\ Phys.\  {\bf 276}, 1 (2007)
  [arXiv:hep-th/0512205].
  %%CITATION = CMPHA,276,1;%%

\bibitem{GLO}
  J.~Gray, A.~Lukas and B.~Ovrut,
  ``Perturbative anti-brane potentials in heterotic M-theory,''
  Phys.\ Rev.\  D {\bf 76}, 066007 (2007)
  [arXiv:hep-th/0701025];\\
  %%CITATION = PHRVA,D76,066007;%%
  J.~Gray, A.~Lukas and B.~Ovrut,
  ``Flux, Gaugino Condensation and Anti-Branes in Heterotic M-theory,''
  Phys.\ Rev.\  D {\bf 76}, 126012 (2007)
  [arXiv:0709.2914 [hep-th]].
  %%CITATION = PHRVA,D76,126012;%%




\bibitem{ISS}
  K.~Intriligator, N.~Seiberg and D.~Shih,
  ``Dynamical SUSY breaking in meta-stable vacua,''
  JHEP {\bf 0604}, 021 (2006)
  [arXiv:hep-th/0602239].
  %%CITATION = JHEPA,0604,021;%%

\bibitem{KKLT}
  S.~Kachru, R.~Kallosh, A.~Linde and S.~P.~Trivedi,
  ``De Sitter vacua in string theory,''
  Phys.\ Rev.\  D {\bf 68}, 046005 (2003)
  [arXiv:hep-th/0301240].
  %%CITATION = PHRVA,D68,046005;%%

\bibitem{OS}
K. Oguiso and T. Shioda, ``The Mordell-Weil lattice of a rational elliptic surface," Comment. Math. Univ. St. Paul. {\bf 40}, 83--99 (1991).

\bibitem{Sc}
C.~Schoen, ``On fiber products of rational elliptic surfaces with section," Mathematische Zeitschrift, Vol. 197(2), 177--199 (1988).

\bibitem{Sh}
T. Shioda, ``On the Mordell-Weil Lattices," Comment. Math. Univ. St. Paul. {\bf 39}, 211--240 (1990).

\bibitem{We}
T.~Weigand, ``Compactifications of the heterotic string with unitary bundles," Fortschr. Phys. {\bf 54}, No. 11, 963--1077 (2006).

\end{thebibliography}
\end{document}